\documentstyle[aas2pp4]{article}

\newcommand{\etal}{et al.}

\begin{document}



\title{Off-Center Collisions between Clusters of Galaxies}

\author{P. M. Ricker}
\affil{Department of Astronomy, University of Virginia, Charlottesville, VA 22903}



\begin{abstract}

We present numerical simulations of off-center collisions between galaxy
clusters made using a new hydrodynamical code based on the piecewise-parabolic
method (PPM) and an isolated multigrid potential solver. The current
simulations follow only the intracluster gas.
We have performed three high-resolution ($256\times 128^2$)
simulations of collisions between equal-mass clusters using a nonuniform grid
with different values of the impact parameter (0, 5, and 10
times the cluster core radius).
Using these simulations we have studied the variation in equilibration time,
luminosity enhancement during the collision, and structure of the
merger remnant with varying impact parameter.

We find that in off-center collisions the cluster cores (the inner regions
where the pressure exceeds the ram pressure) behave quite differently from
the clusters' outer regions.
A strong, roughly ellipsoidal shock front, similar to that noted in previous
simulations of head-on collisions, enables
the cores to become bound to each other by dissipating their kinetic energy
as heat in the surrounding gas.
These cores survive well into the collision,
dissipating their orbital angular momentum via spiral bow shocks. 
After the ellipsoidal shock has passed well outside
the interaction region, the material left in its wake falls back onto
the merger remnant formed through the inspiral of the cluster cores, creating
a roughly spherical accretion shock. For less than one-half
of a sound crossing time after the cores first interact the total X-ray
luminosity increases by a large factor; the magnitude of this increase
depends sensitively on the size of the impact parameter.

Observational evidence of the ongoing collision, in the form of bimodality
and distortion in projected X-ray surface brightness and temperature maps, is
present for 1--2 sound crossing times after the collision, but
only for special viewing angles. The remnant
actually requires at least five crossing times to reach virial equilibrium.
Since the sound crossing time can be as large as 1--2 Gyr, the equilibration
time can thus be a substantial fraction of the age of the universe.
The final merger remnant is very similar for impact parameters of zero and five
core radii. It possesses a roughly isothermal core, with central density and
temperature twice the initial values for the colliding clusters. Outside the
core the temperature drops as $r^{-1}$, and the density roughly as $r^{-3.8}$.
The core radius shows a small increase due to shock heating during the merger.
For an impact parameter of ten core radii the core of the remnant possesses
a more flattened density profile, with a steeper dropoff outside the core.
In both off-center cases the merger remnant rotates, but only for the
ten-core-radius case does this appear to have an effect on the structure of
the remnant.

\end{abstract}




\clearpage


\section{Introduction}

Galaxy clusters are the largest bound structures in the present-day
universe. However, they are not the oldest.
For galaxy clusters the notion of `age' does not necessarily
refer to the time elapsed since a specific formation event occurred,
after which they remained either static in their properties or very
slowly evolving. Indeed, a growing body of observational and theoretical
evidence indicates that for many clusters `formation' means an ongoing
sequence of mergers and interactions with other clusters.
Understanding this process is important not only from the
standpoint of understanding cluster evolution, but also because of its
implications for cosmology.


Recent ROSAT and ASCA observations of rich clusters of galaxies have
shown that many of them are either in the process of merging with
another cluster or have recently undergone such a merger.
The ROSAT results, because of poor spectral
resolution but good spatial resolution, emphasize the variation
in gas density and hence the depth of the potential well in a cluster.
These
include observations of the Coma cluster (\cite{WBH93}),
Abell 2256 (\cite{Bri91}), and Abell 2163 (\cite{EAB95}).
The signatures of merging events in these observations include
distorted X-ray isophotes (presumably elongated along the
collision axes); offsets of the gas centroid, as indicated by the X-ray
emission, and the center of mass of the galaxy distribution; and bimodal
distributions of emission.
The ROSAT results are consistent with
earlier results from Einstein observations which suggested that
at least 22\% of clusters show significant X-ray substructure
(\cite{ForJon94}).
In the more recent ASCA observations of
Abell 754 (\cite{HenMar96}), which emphasize the variation in gas temperature
through its influence on the X-ray spectrum, the presence of a ridge
of high-temperature emission has been taken as indicative of an
ongoing merger event. However, significant temperature variations have not
been seen by ASCA
in other merger candidates identified from ROSAT data, including
Abell 2256 and Abell 2163 (\cite{Mar96}).


Although we do not yet completely understand the
physical mechanisms behind and
cosmological implications of galaxy cluster mergers,
it is not surprising that they should occur frequently.
In hierarchical models of large-scale structure formation, objects like
galaxy clusters form through the merging and then accretion of smaller
clumps of matter. In these models the present-day fraction
of clusters which have significant substructure
can give us constraints on the value of the density parameter $\Omega$
(\cite{RLT92}), and the mass density profiles resulting from mergers
provide information about the spectrum of primordial density fluctuations
(\cite{HofSha85}). However, although analytical approaches based on the
spherical top-hat model and the Press-Schechter (1974) mass function have
provided much insight into these implications, the complex physical
interactions present during a cluster merger can only be
studied adequately through numerical simulation.


To date simulations of galaxy cluster evolution (either
$N$-body or mesh-based) have typically followed one of three approaches.
Some simulations have focused on the collapse and relaxation of a single
overdense region which has been prepared as a `typical' example given
a large-scale structure model (e.~g.\ \cite{Evr90}; \cite{EMN96}).
Others have produced
clusters within large-scale structure simulations, then followed their
evolution separately from the rest with greater spatial resolution
(e.~g.\ \cite{KatWhi93}). Both types of simulation produce realistic
clusters in an environment which permits comparison of large-scale
structure theories but which renders analysis of individual merging
events difficult because of the uncontrolled initial conditions
for each event and the frequent occurrence of multiple simultaneous mergers.

A third type of cluster evolution simulation has focused on
collisions of idealized clusters or subclusters
(e.~g.\ \cite{RBL93}; \cite{SchMul93}; \cite{PTC94}; \cite{RSM97})
to gain insight into the merger process without
complicating the problem with multiple mergers.
This approach also permits mergers to be studied with better resolution
than is possible in large-scale structure simulations.

Roettiger \etal\ (1993)
simulated the head-on collision of two King-model clusters of unequal
mass using the ZEUS-3D finite-difference code with a nonuniform grid
for the gas dynamics and the Hernquist (1987) treecode for the dark matter.
The resolution of these simulations was such that the smallest zones
were one-half the core radius of the smaller cluster.
During the collision they observed a double-peaked X-ray emission profile
and a bar of X-ray emission perpendicular to the collision axis which
resulted from a strong central shock (Mach number $\sim 4$). They also
observed a high-velocity ($\sim 2800$ km/s), ordered flow of gas through
the center of the dominant cluster which they suggested might disrupt
cooling flows and be responsible for the bending of radio jets in
cluster galaxies. Although their simulation did not include cooling,
a later simulation by Burns \etal\ (1994) of a collision between the Coma
cluster and the NGC 4839 group, using the same code but including cooling and
having somewhat higher core resolution, obtained similar results.

Schindler and M\"uller (1993) also studied head-on collisions derived from slightly
less controlled initial conditions using the piecewise-parabolic method
(PPM) for the gas dynamics and Aarseth's (1972) code for collisionless
galaxy halos. During the course of their collisions they observed
multiple interacting shocks which led to X-ray luminosity enhancements of
about a factor of 1.5 and temperature enhancements of a factor of 5,
with a non-isothermal final state as the result. These calculations
used a uniform grid with about 2.5 cells per core radius.

Neither Roettiger \etal\ (1993),
Burns \etal, nor Schindler and M\"uller included the baryonic
contribution to the gravitational potential, thus limiting their conclusions to
gas-poor clusters (less than 5\%--10\% of the total mass). However, some
clusters can have as much as 30\% (for a Hubble constant of 
50 km s$^{-1}$ Mpc$^{-1}$) of their total mass in the intracluster
medium (\cite{DJF95}; \cite{WhiFab95}).
The virial estimates of total cluster masses on which
this ROSAT finding is based are supported by recent simulation work by
Evrard \etal\ (1996). The gas in these clusters may therefore have a significant
effect on the potential, rendering a self-consistent treatment necessary.

Pearce \etal\ (1994) performed head-on collision simulations
of equal-mass clusters using an adaptive
smoothed-particle hydrodynamics (SPH) code which does include both
gas and dark matter in the potential. They found that the gas, which
made up 1/8 of their clusters' total mass, tends to
produce an extended constant-density core after the collision,
whereas the dark matter component does not. However, while they note
the presence of shocks due to the collision of the gas components of the
clusters, they do not discuss the observational appearance of the
merging system in detail or compare it to the finite-difference results.
More recently,
Roettiger \etal\ (1997) have used a PPM/particle-mesh code
to simulate a collision intended to reproduce the observations of A754, varying
collision parameters at low resolution and then resimulating with the most promising
parameter values at high resolution.


With the exception of Roettiger \etal\ (1997),
none of these authors have considered off-center collisions in detail,
although Pearce \etal\ mention that preliminary results from such simulations
indicate that the gas receives additional support from rotation in the
merger remnant. Interest in off-center collisions is growing, however, as
ASCA temperature maps of clusters such as A754 (\cite{HenMar96}) begin to
give us more detailed information about off-center cluster mergers in progress.
The net angular momentum imparted to a cluster by many such
collisions is probably small, since the angular momenta brought by collisions
in different directions will tend to cancel each other out. Also, the net
spins induced in collapsed objects by tidal torques due to inhomogeneities in
the large-scale mass distribution have been shown to be small (\cite{Pee93}).
However, during their pre-collision evolution, colliding subclusters should
experience some deflection due to distant collisions with other subclusters.
Given the great difference in behavior to be expected between the limiting
cases of a simple distant deflection and a head-on collision, we would do well
to consider the dynamics of intermediate cases and its effects on the long-term
evolution and observable properties of clusters.

In order to study galaxy cluster evolution, and in particular off-center
collisions of galaxy clusters, we have developed a parallel
hydrodynamics code based on PPM which self-consistently solves for the
gravitational potential of the gas using a multigrid algorithm with isolated
boundary conditions.
The code uses a nonuniform grid, permitting it to resolve the inner
regions of the colliding clusters while simultaneously tracking the
development of the shocks which result from the collision.
We have added radiative cooling and linked the PPM code to a particle-mesh
$N$-body solver for dark matter (written by Scott Dodelson), but the
simulations described in this paper include only nonradiating gas.
While these gas-only calculations necessarily exclude the substantial
contribution of collisionless effects to cluster evolution, they nevertheless
represent an important benchmark for more realistic calculations including
dark matter, because it is far more straightforward to understand both
the physical and numerical properties of the hydrodynamical results in
isolation. This is analogous to the approach taken prior to the advent of
combined $N$-body/hydro codes, in which the collisionless dynamics of
the dark matter was studied in isolation using $N$-body codes with
well-understood properties. In a forthcoming paper we will discuss simulations
performed with our new hybrid code, using the results presented here as a
basis of comparison.

The paper is organized as follows. Section 2 gives a brief description
of the code and the initial and boundary conditions used; a more complete
description, including the equations used and the results of running the
code on several standard test problems, will be presented in another
forthcoming paper. Section 3 describes
a resolution study performed using a single cluster at rest at the
center of the grid. Section 4 describes results from three high-resolution
simulations of collisions between equal-mass clusters at different
impact parameters. Section 5 discusses the equilibration times and X-ray
brightening observed in these calculations, as well as the thermal and
rotational properties of the merger remnants. Section 6 summarizes our
conclusions.


\section{Numerical methods}


\subsection{Hydrodynamics}

To solve the equations of hydrodynamics for the cluster gas we use the
direct Eulerian formulation of the piecewise-parabolic method
(PPM; \cite{ColWoo84}) on a nonuniform Cartesian grid. In contrast
to most numerical studies of cluster evolution which use smoothed particle
hydrodynamics (SPH) (e.~g.\ \cite{Evr90}) for its ability to
dynamically adjust spatial resolution,
simulations using PPM offer better shock handling for comparable
spatial resolution, since they require very little artificial viscosity.
SPH resolves shocks only poorly and requires substantial amounts of
artificial viscosity.
Furthermore, in low-density regions where SPH represents the gas with
relatively few particles, PPM obtains more accurate temperatures 
(\cite{Kan94}). The error properties of finite-volume methods in
general, of which PPM is a specific example, are better understood than
those of SPH.
For these reasons we believe PPM to be the more appropriate method at present
for studying the evolution of the collisional component of galaxy clusters.

Others have used PPM to study cluster evolution.
Schindler and M\"uller (1993)
used PPM together with the $N$-body solver of Aarseth (1972)
to evolve the gas in their simulation on a $60^3$ uniform mesh covering
little more than the two colliding clusters they studied.
Bryan \etal\ (1994) used PPM on a much larger mesh with $270^3$ cells to study the
evolution of many poorly resolved clusters over a volume large enough to
contain a statistical sample of clusters.
Roettiger \etal\ (1993) used a non-Godunov-based finite-difference method
similar to PPM in its use of parabolic interpolants to follow two merging
clusters; their grid contained $187\times 65^2$ cells and had nonuniform
spacing which permitted good resolution of the cluster cores.
More recently, Roettiger \etal\ (1997) have used a PPM/particle-mesh hybrid code
to simulate the evolution of Abell 754 on a $512\times256^2$ grid with 
a maximum resolution of 1/10 the cluster core radius.

However, with the exception of Bryan \etal\ and Roettiger \etal\ (1997),
none of these authors self-consistently solved for the contribution
of the gas to the potential. This is a serious shortcoming, one not shared by
the SPH codes; even a gas mass fraction of 10\% can significantly affect the
potential on small scales if the gas and dark matter distributions are not
the same. In our
code we therefore include the gas density in the computation of the potential,
which is described in the next subsection.
Although the results presented here do not include dark matter, we have
linked our hydrodynamical code to an $N$-body code based on the particle-mesh
method, with the two codes sharing the same potential solver. We plan to
report on results obtained with the new hybrid code in the near future.
The new code also includes radiative cooling, which we do not include in the
results presented here.

In order to look for observational signatures of recent on- or
off-center merger events, one must be able to study a computational volume
which is large enough to contain the shocks produced by the merger until they
cease to be important.
Furthermore,
our single-cluster test runs, described in the next section, suggest that
to do the problem correctly one must use a grid spacing in the inner regions
of the clusters of at most one-half the core radius $r_c$.
In order to begin to reconcile these two requirements
we have used for our high-resolution runs a nonuniform
grid of $256\times 128^2$ zones.
The innermost $N_{x,{\rm inner}}=192$ zones of the $x$-axis have uniform
widths equal to $0.28r_c$.
The innermost $N_{yz,{\rm inner}}=72$ zones in $y$ and $z$ have
uniform widths equal to $0.39r_c$.
Outside the uniformly gridded region in each coordinate direction
the zone widths increase at the rate of 5\% per zone as one moves away from
the center of the grid.
The size of the uniformly gridded region was chosen to maximize resolution
and minimize any effects due to grid nonuniformity in the region where most
of the complex shocks form during the simulation.
In terms of the initial cluster radius $R=9r_c$, our computational grid has
physical dimensions $11R\times 8.1R\times 8.1R$.

In test calculations of the Sedov-Taylor explosion problem (\cite{Sed59})
we have found that using first-order operator
splitting to create a three-dimensional PPM scheme does not adequately
maintain the 90-degree rotational symmetry of a uniform grid, particularly
in the central part of the grid where a rarefied region develops
in the Sedov-Taylor problem. We therefore
use a second-order splitting, generalized from Strang (1968) to three
dimensions. Although this doubles the cost of a hydrodynamical timestep
relative to first-order splitting, the spherically symmetric, centrally
concentrated nature of our initial density and pressure field makes it
important to preserve this symmetry.

In a realistic simulation of cluster evolution, tides from mass concentrations
outside of the grid would affect the simulated objects. Mass inflow through
the grid boundaries would also be expected. However, in our simulations we
are studying the behavior of a single merging system, so vacuum boundaries are
appropriate. To implement these we use the standard zero-gradient boundaries
with a slight difference. If purely zero-gradient boundaries are used, any
inward-directed velocity in the interior zones adjacent to the boundary will
result in a completely artificial inward mass flux. Since this mass
flux adds to the depth of the potential, adding to the inward acceleration
in these zones, this problem can destabilize the entire calculation.
To prevent such artificial inward flows we modify the zero-gradient
prescription when velocities in the interior zones adjacent to the boundaries
are directed inward. For these zones we apply Dirichlet boundary conditions
to the component of velocity normal to the boundary. All other variables,
including velocity components parallel to the boundary, are given zero-gradient
boundary conditions as usual. This suffices for the interpolation step
in PPM; for the flux computation step we set the boundary flux of all
conserved quantities (mass, momentum, and energy) to zero if it would otherwise
be directed inward.


\subsection{Gravitation}

At each timestep of the hydrodynamical code we solve the Poisson equation for
the gravitational potential using a full multigrid/nested iteration code
(\cite{Bra77}; \cite{Hac85}). This method accelerates the convergence of
standard elliptic relaxation techniques by iteratively solving on grids of
increasing coarseness.
Because long-wavelength errors are responsible for the
slow convergence of standard iteration techniques, solving on coarse grids
accelerates convergence by turning long-wavelength modes into short-wavelength
modes. The resulting speedup is sufficient to make multigrid methods 
competitive with direct methods, such as those based on the fast Fourier 
transform, even for nonlinear equations or unusual boundary conditions. 

In our case the primary
reasons for using a multigrid method are the need for ease and portability of
parallelization and the nature of the boundary
conditions for our problem. FFT-based Poisson solvers
are most efficient when working with periodic, Dirichlet, or Neumann
boundaries on a uniform grid. However, in the colliding cluster problem,
it is more natural to permit the potential to
go to zero at infinity. Methods exist to handle these `isolated'
boundaries using direct solvers (e.~g.\ \cite{Jam77})
but suffer from the disadvantages of the direct methods on which they are
based. To handle isolated boundaries with multigrid we have developed a
method similar to that of James. We first obtain a solution with Dirichlet
boundaries, then use this to compute image charges on the boundaries under the
assumption that the potential is zero not only on the boundary but everywhere
outside it as well. We then compute the isolated potential of the image charges
and subtract it from the Dirichlet solution.
To calculate the isolated potential of the image charges, we use a truncated
spherical harmonic expansion of the image charge distribution about its center
of mass to estimate boundary values for the potential, then solve Laplace's
equation in the interior using these boundary values.
Although this method requires two calls to the multigrid solver, we only need
to compute image moments and boundary values for points on the boundary.
We have tested this potential solver with mass distributions drawn from
collision runs performed on a nonuniform $128\times 64^2$ grid similar to that
described in the previous section, and we find that 
an acceptable compromise between computation time and accuracy is reached
if one truncates the multipole expansion at $\ell = 10$. The typical
change in the potential $\phi$ in going to larger $\ell$ is less than 1\%.


\subsection{Initial conditions}

We are interested in gaining insight into the physics of
cluster collisions without the complication of initial conditions
drawn from a large-scale structure simulation.
By examining the parametric dependence of the observed features of such
events we can later check cosmological theories by determining what
distribution of collision parameters they produce. Accordingly we have chosen
for our initial conditions nonsingular isothermal spheres, approximated by
the modified Hubble law,
\begin{equation}
\rho(r) = {\rho_0\over \left[{1+\left({r/r_c}\right)^2}\right]^{3/2}} ,
\label{Eqn:HubbleLaw}
\end{equation}
\noindent where $\rho_0$ is the central mass density
and $r_c$ is the cluster core
radius, as in the paper by Pearce \etal\ (1994).
However, instead of truncating the distribution (\ref{Eqn:HubbleLaw})
beyond the radius $R=16r_c$ as they did, we have used the smaller value
of $9r_c$. The exact value of $R/r_c$, if it is of order 10 or so, is not
important for the dynamics of the inner regions of the clusters, but it
does affect our ability to resolve the cluster cores properly.
We assume that the density perturbations which created the
clusters have entered the strongly nonlinear regime well before the
start of the simulation, so we use a background spacetime which is flat
and nonexpanding.
Since PPM is a finite-volume scheme which uses
cell-averaged quantities (in contrast to a finite-difference scheme, which
manipulates the values of fields at grid points), we set
the density in each cell using Monte Carlo averaging of the density
profile (\ref{Eqn:HubbleLaw}).
The temperature is specified by the core radius and central number
density via the standard definition of the core radius (\cite{BinTre87}):
\begin{equation}
{kT_{\rm init}\over\mu} = {4\over 9}\pi G\rho_0 r_c^2\ .
\label{Eqn:CoreRadiusDefn}
\end{equation}
\noindent We set the pressure in each cell by Monte Carlo-averaging $\rho(r)T$
together with the density. To convert temperature to pressure we use
\begin{equation}
p = {\rho kT_{\rm init}\over\mu}\ ,
\label{Eqn:Eqn of State}
\end{equation}
\noindent where $k$ is Boltzmann's constant.
Here $\mu$ is different from the average atomic mass
$\langle m\rangle$ because of the contribution of electrons to the gas
pressure; assuming a
primordial composition (25\% helium, 75\% hydrogen by mass), we have
$\mu \approx 0.59m_{\rm H}$ and $\langle m\rangle \approx 1.2m_{\rm H}$.
As our gas is
hot enough to be completely ionized, we use $\gamma=5/3$ as our ratio of
specific heats.

\begin{figure}
  \figurenum{1}
  \plotone{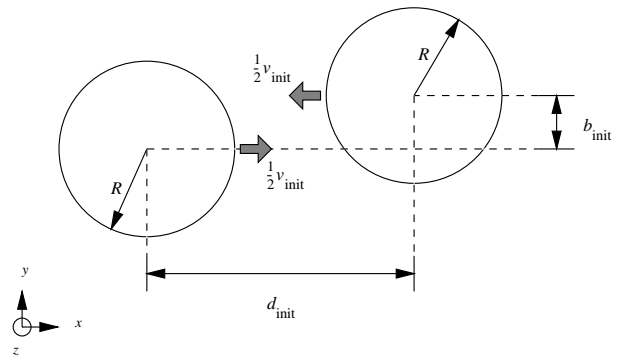}
  \caption{Initial conditions for cluster collision runs.}
  \label{Fig:Initial conditions}
\end{figure}

In the simulations presented here, both clusters have the same central
density and core radius, hence the same mass and temperature.
We use units in which each cluster initially has
unit mass, unit radius, and unit temperature.
The time unit is one sound crossing time,
\begin{equation}
t_{\rm sc} \equiv {R\over c_s} = 9r_c\sqrt{\mu\over\gamma kT_{\rm init}}\ ,
\end{equation}
\noindent so in our units the initial sound speed $c_s$ is also unity.
In these units the central density $\rho_0$ is approximately 30.5, and
the core radius $r_c$ is 1/9. The scaling of our simulations to physical
units is determined by our choices for the values of $m_{\rm H}/k$ and
Newton's gravitational constant $G$; we use $m_{\rm H}/k \approx 2.86
t_{\rm sc}^2T_{\rm init}R^{-2}$ and $G \approx 1.14 R^3 M^{-1} t_{\rm sc}^{-2}$.
Outside the two clusters, the density is set to $10^{-8}\rho(R)$, and
the temperature is the same as it is inside the clusters.

The two clusters are initialized with centers in the $xy$-plane,
separated by a distance $b_{\rm init}$ (the impact parameter) in the
$y$-direction and a distance $d_{\rm init}$ in the $x$-direction
(Figure \ref{Fig:Initial conditions}).
For the calculations described here
we performed three $256\times 128\times 128$ runs using different values of the
impact parameter $b_{\rm init}$.
In run A, $b_{\rm init}=0$; in run B, $b_{\rm init}=5r_c$; and in run C,
$b_{\rm init}=10r_c$.
The value of $d_{\rm init}$ was held fixed at $4R$ for all of the runs,
so that the clusters started close enough together to collide within
a short time but far enough apart so that any numerical artifacts
due to the initial conditions would be smoothed out before the cores
interacted.
The initial relative velocity, $v_{\rm init}$,
was chosen to be approximately equal to
the free-fall velocity for the two clusters at a distance $d_{\rm init}$,
and was directed along the $x$-axis.
All three runs used the same value for $v_{\rm init}$ (unity).
The results of these runs are discussed in Sections 4 and 5.


\section{Single-cluster tests}

To study the behavior of our code on a simple yet realistic test problem,
we simulated a single cluster at the center of a nonuniform grid similar to
that used in our collision simulations. Since the gas density profile is
approximately that of an isothermal sphere, these initial conditions should
produce an approximately static configuration. We performed three runs at
grid sizes of $32^3$ (run T1), $64^3$ (run T2), and $128^3$ (run T3) using
nonuniform gridding in all three dimensions. The physical box size was kept
constant at $L \approx 6R$, and the
grid nonuniformity, described in the previous section, was kept
constant at $5$\%. This leaves the number of zones in the uniform
inner region, $N_{\rm inner}$, and the zone size in this region, $\Delta$, as
free parameters. The values of these parameters used in the test runs are
summarized in Table \ref{Table:Single Cluster Grid Parameters}.
In each run the central density $\rho_0$ and core radius
$r_c$ were given the same values used in the collision runs.
Varying the resolution in this way permitted us to determine how
core resolution affects the accuracy of the solution in the central regions
of the cluster, and also to control for the deviation of our initial conditions
from true isothermal spheres in the collision runs.

\begin{table*}
\tablenum{1}
\caption{Grid parameters used in the single-cluster test runs.
  \label{Table:Single Cluster Grid Parameters}
  }
\begin{center}
\begin{tabular}{llll}
Run     & Grid size       & $N_{\rm inner}$       & $\Delta/r_c$ \\
\tableline
T1      & $32^3$          & 16                    & 1.36  \\
T2      & $64^3$          & 32                    & 0.582 \\
T3      & $128^3$         & 64                    & 0.212 \\
\tableline
\end{tabular}
\end{center}
\end{table*}

\begin{figure}
  \figurenum{2}
  \plotone{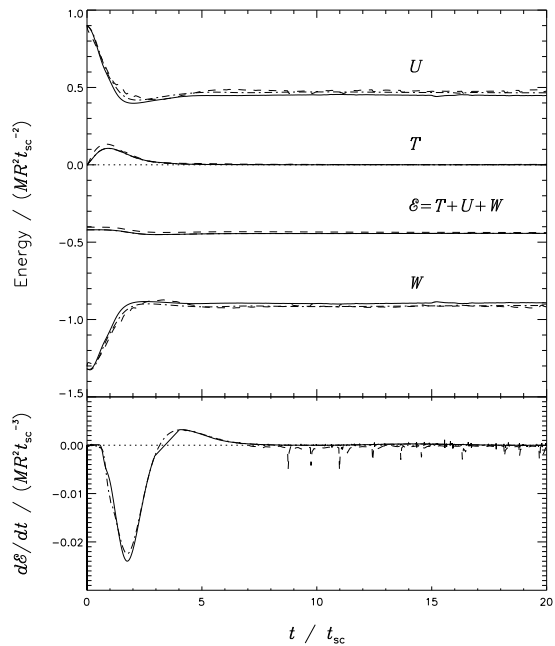}
  \caption{The upper plot shows the kinetic, internal, and potential energies
  in the single cluster tests. The total energy is also plotted. The lower plot
  shows the time derivative of the total energy. Dashed lines indicate run
  T1 ($32^3$), dot-dashed lines, run T2 ($64^3$), and solid lines, run T3
  ($128^3$).}
  \label{Fig:Single Cluster Energy Plot}
\end{figure}

Figure \ref{Fig:Single Cluster Energy Plot} presents the average kinetic
($T$), internal ($U$), and potential ($W$) energies of the cluster in each
of the three test runs.
For each run the total energy (${\cal E}=T+U+W$),
less the amount lost from the grid between $t=t_{\rm sc}$
and $t=3t_{\rm sc}$,
is constant to better than 1\%, with the energy differing slightly between
the runs because of slight differences in the size of the box used.
However, in each run during the first 2--3 sound crossing
times the internal and potential energies decrease in magnitude, and the
kinetic energy briefly increases from zero. This occurs because
our initial conditions contain an artificial shock where the Hubble-law
profile is truncated. To keep the near vacuum outside the clusters
in pressure equilibrium with the cluster, we would have to set the
temperature in these regions to a very large value, forcing the timestep
to be unacceptably small (via the Courant condition). We therefore allow a
pressure gradient to exist
at the edges of the cluster. Initially this gradient drives a rapid expansion
of the `vacuum material' outside the cluster. The cluster itself also expands,
cooling by about a factor of two and driving about 15\% of the total mass 
off the grid between $t=t_{\rm sc}$ and $t=3t_{\rm sc}$.
As the pressure gradient is relieved by this expansion, the expansion slows and
stops. The material remaining on the grid briefly falls back toward the
center of the cluster before reaching hydrostatic equilibrium.
Because the computational box is not much larger than the cluster itself,
at late times the density distribution in the `vacuum' regions is not quite
spherical.
We use a larger box in the collision simulations to avoid this effect.

The time required for the cluster to reach virial equilibrium, as determined
using the method described in Section 5, was approximately $(2$--$3)t_{\rm sc}$
in each test run. In order to control for this relaxation in the collision runs
we place the cluster centers far enough apart initially that their cores
do not interact until this amount of time has passed.

\begin{figure}[tbh]
  \figurenum{3}
  \plotone{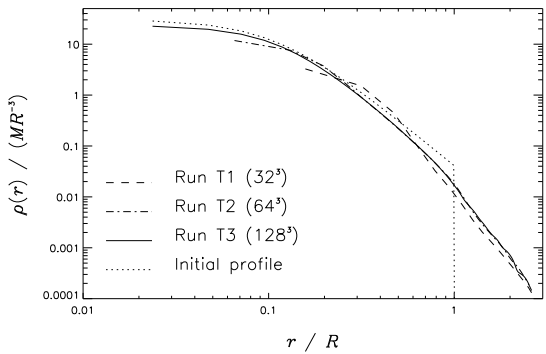}
  \caption{Angle-averaged density profiles in
  single-cluster test runs at $t = 20t_{\rm sc}$. The dotted
  line indicates the initial truncated Hubble-law profile, the dashed
  line indicates the final profile for run T1, the dot-dashed line indicates
  run T2, and the solid line indicates run T3.}
  \label{Fig:Single Cluster Density Profiles}
\end{figure}

\begin{figure}[tbh]
  \figurenum{4}
  \plotone{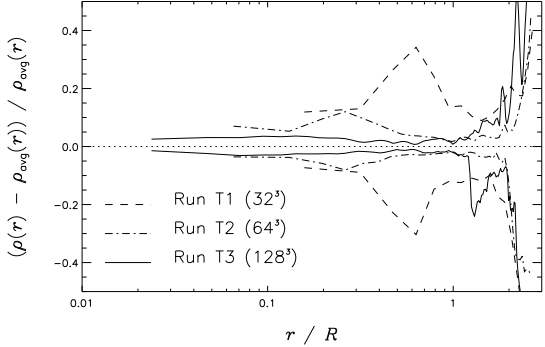}
  \caption{Fractional deviation from the average of the minimum and maximum
  density as a function of radius in the single-cluster test runs at
  $t=20t_{\rm sc}$.}
  \label{Fig:Single Cluster Density Asymmetry}
\end{figure}

Figure \ref{Fig:Single
Cluster Density Profiles} shows angle-averaged density profiles for the
three test runs at $t=20t_{\rm sc}$, well after each has reached a steady state.
We calculate each average profile by linearly interpolating from the
computational grid along 1000 randomly chosen lines emanating
from the center of mass; each line is uniformly sampled in radius out to the
closest external grid boundary.
The percentage deviation from average of the minimum and maximum density as
a function of radius is plotted for each run in Figure
\ref{Fig:Single Cluster Density Asymmetry}.
In run T1, for which the central zone size was somewhat larger than $r_c$,
the final profile differs substantially from the initial profile, even in the
outer regions of the cluster ($2r_c < r < R$), where the density is almost
a factor of two larger than in the highest-resolution run. The density profile
outside the cluster has not even converged, although its logarithmic
slope has.
In run T2, which used a central zone size of about $0.6r_c$, the outer regions
of the cluster have converged to the high-resolution result, but the density at
$r\approx 2r_c$ is overestimated by a factor of 1.2,
and the region inside one core radius is low by a factor of 1.7.
Although this resolution does not allow study of scales smaller
than $r_c$, it appears to be needed in order for the outer regions of the
cluster to be correct. In both run T1 and run T2 the effect of poor central
resolution shows up as an overestimate of the density in the zones just
outside the center and an underestimate in the central zones. This effect
is most
likely due to errors in the central pressure and potential gradients arising
from the use of a Cartesian grid, as can be seen in Figure
\ref{Fig:Single Cluster Density Asymmetry}. The density asymmetry in runs
T1 and T2 reaches its maximum in the zones just outside the center; for run T1
this maximum is about 35\%, and for run T2 it is about 12\%. Outside the
cluster the asymmetry increases to more than 50\% in each of the runs because
of the finite box size effects mentioned earlier.

The density profile in run T3, which used a zone size equal to one-fifth of a
core radius, agrees with the $64^3$ result outside $r\approx 2r_c$, but
increases smoothly and monotonically to its central value, which is 20--30\%
smaller than
the initial central value. This decrease in the central density is to be
expected because of the initial pressure-driven
expansion of the cluster, the consequent loss of 15\% of the initial mass
through the grid boundary, and the lack of a radiative cooling mechanism.
For $2r_c\le r\le 8r_c$ the density drops as $r^{-3.1}$, slightly more
steeply than the initial $r^{-3}$ profile. Outside of the cluster the density
drops as $r^{-4.8}$.
The density asymmetry inside the cluster is everywhere less than 5\%, and
in the outer regions of the cluster it drops as low as 1\%.
Because of the average density profile's agreement with the $64^3$ result, its
smoothness and monotonicity, and its spherical symmetry,
we regard the $128^3$ result as being adequately converged. The central zone
size used in
our collision runs is approximately one-third of a core radius, between the
zone sizes in runs T2 and T3.

\begin{figure}[tbh]
  \figurenum{5}
  \plotone{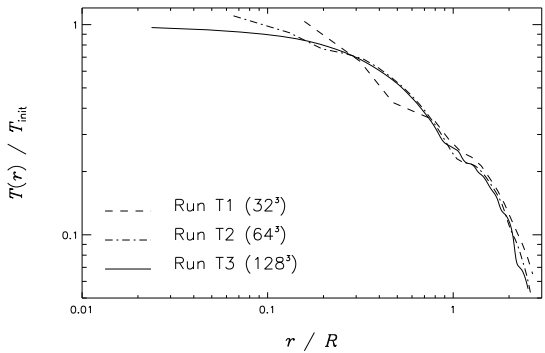}
  \caption{Angle-averaged temperature profiles in single-cluster test runs at
  $t = 20t_{\rm sc}$.}
  \label{Fig:Single Cluster Temperature Profiles}
\end{figure}

In none of the test runs does the cluster remain isothermal.
During the cluster's initial expansion, most of the adiabatic cooling
occurs in its outer regions.
Figure \ref{Fig:Single Cluster Temperature Profiles} shows the angle-averaged
temperature profile in each run at $t=20t_{\rm sc}$. The temperature at the
edge of the cluster ($r=R$) has fallen by a factor of five in each run, and
at the edge of the grid it is smaller again by a factor of two. The central
temperature is close to its original value of unity in each case, but in runs
T1 and T2 it is about 5--10\% higher. This is consistent with the underestimated
central density and resultant shallowness of the density profile near the
center in these runs. In run T1 the temperature in the outer regions
of the cluster is badly underestimated.
As with the density profile, the temperature
profile in run T3 increases smoothly and monotonically to its central value,
0.98. Outside of $r\approx 3r_c$ it agrees well with the result from run T2.

We conclude from these observations
that resolving the core region is of great
importance for minimizing numerical effects due to the Cartesian geometry
of our grid. These effects include an underestimate of the central density
relative to the regions just outside the center, asymmetry in the density
profile just outside the center, and excessive heating of the central zones.
The amount of material lost from the cluster during its expansion, and the
slope of the density and temperature profiles outside the cluster, do not
appear to depend significantly on the core resolution, but poor core
resolution does cause an overestimate of the density and a substantial
underestimate of the temperature in the regions just outside the core.
A zone size smaller than $r_c/2$ for the
region $r<3r_c$ appears to be needed in order to avoid these effects.

Although we regard this resolution as adequate for the study of our
nonradiating cluster gas, when comparing our results to real systems the
bremsstrahlung cooling time must be kept in mind.
For radii smaller than $r_c$ the initial radiative cooling timescale,
\begin{equation}
t_{\rm cool} = 2.6\left({R\over{\rm Mpc}}\right)
		  \left({\rho\over MR^{-3}}\right)^{-1}
		  \left({T\over T_{\rm init}}\right)^{1/2} t_{\rm sc}\ ,
\end{equation}
\noindent can be comparable to the sound crossing time.
In order to correctly resolve
scales smaller than this it is necessary to include radiative cooling,
which in addition to permitting the study of cooling flows would tend to
stabilize the cluster against the spreading caused by errors in the
pressure and potential gradients.


\section{Results}

In this section we present results from three collision simulations using
equal-mass clusters at different impact parameters: zero, five, and ten
core radii. The three collisions were followed until $t=12t_{\rm sc}$,
well beyond the point at which the merger remnant reached equilibrium
in each case. The different collisions all involve multiple interacting shocks
whose behavior and effect on the overall progress of the collision change
with increasing impact parameter. Here we consider this variation as revealed
by snapshots of the density, temperature, and velocity fields, and by the
projected X-ray surface brightness and temperature fields, at several different
timesteps.


\subsection{Head-on collision}

\begin{figure}
  \figurenum{6a}
  \plotone{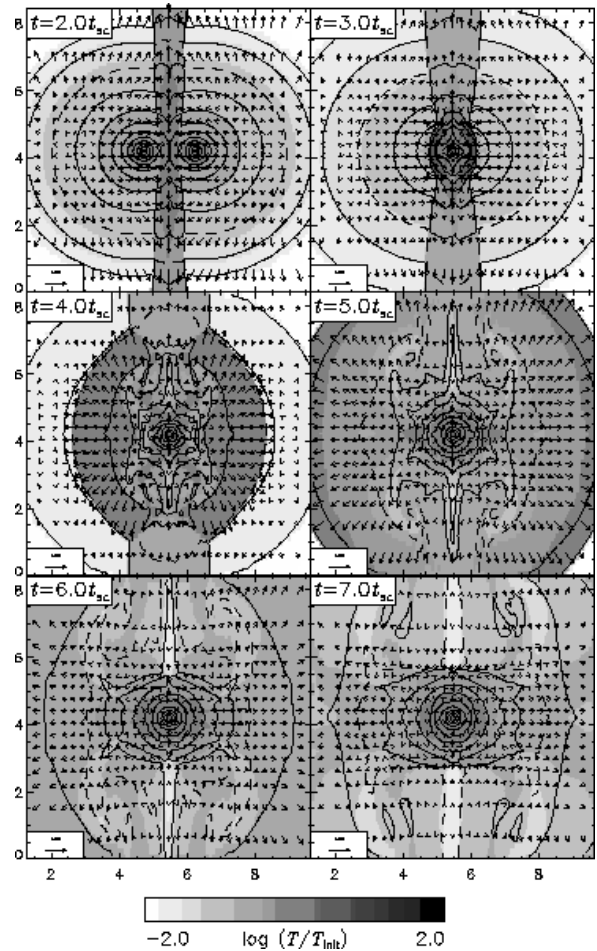}
  \caption{Density and
  temperature snapshots for run A in the $xy$-plane passing through
  the cluster centers. Units are as described in the text, with
  $R = M = T_{\rm init} = t_{\rm sc} = 1$.
  Density contours are separated by a factor of three; the dashed contour
  indicates $\log\rho=-1$.
  Shading indicates the logarithm of the temperature.
  Velocity arrows are drawn for every eighth cell, with the fiducial arrow at
  the corner of each plot representing $v=1$.}
  \label{Fig:Run A Full XY Slice Plots}
\end{figure}

\begin{figure}
  \figurenum{6b}
  \plotone{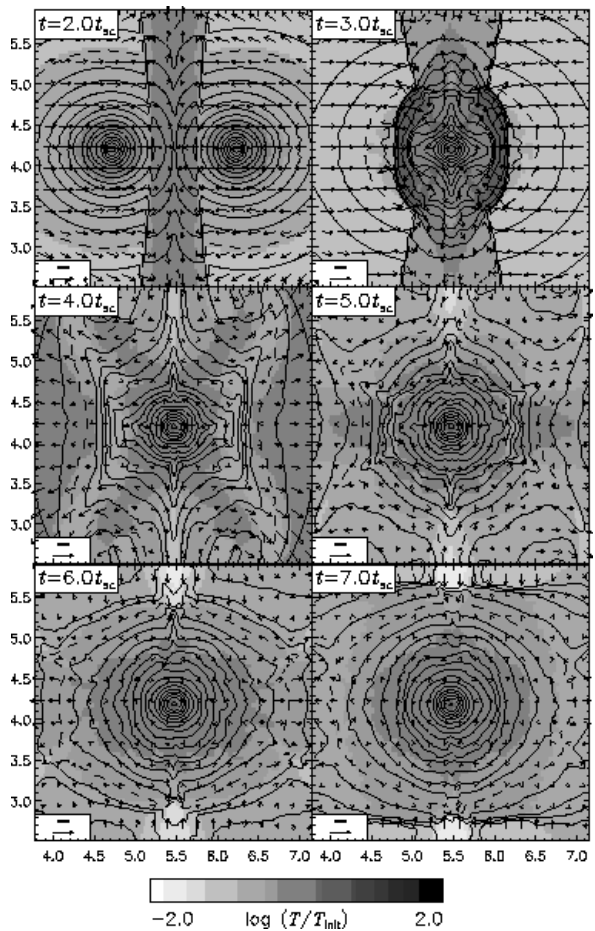}
  \caption{Detail of the innermost 16\% of
  Figure \protect\ref{Fig:Run A Full XY Slice Plots}
  (the $xy$-plane). Log density contours are spaced every 0.25.
  Velocity arrows are drawn for every sixth cell; the fiducial arrows
  indicate $v=0.5$.}
  \label{Fig:Run A Zoom XY Slice Plots}
\end{figure}

The first run we performed, run A, simulated a head-on collision
($b_{\rm init}=0$). In Figures \ref{Fig:Run A Full XY Slice Plots} and
\ref{Fig:Run A Zoom XY Slice Plots} we present snapshots of the density,
temperature, and velocity fields at six representative times during the
simulation in the plane having a constant $z$-value of 1/2 the box height.
The first figure shows an $8.1R\times 8.1R$ section of the complete plane,
and the second shows the
innermost 16\% of this region. In both plots the abscissa is the $x$-axis,
and the ordinate is the $y$-axis.

These figures show several important features of the collision. Of
particular interest is the contrast between the behavior of the
cores and that of the outer regions of each cluster. In the
first snapshot of each figure, corresponding to $t=2t_{\rm sc}$,
the outer regions of the clusters have
collided, forming a pair of slow-moving shock waves which propagate along
the collision axis in both directions. Because of the clusters' gravitational
potential,
these shocks move more rapidly in the regions away from the collision axis,
giving them a pinched appearance at the center of the grid. In the region
between the shocks, the temperature is higher than its pre-shock value
by about a factor of three. The approaching cluster cores
compress this shock-heated gas, causing it to be expelled from the collision
axis at about one-half the clusters' initial free-fall velocity.
The speed of this material increases by a factor of two as it accelerates
down the pressure gradient.

As the cluster cores collide at the center of the grid just before
$t=3t_{\rm sc}$, they produce a region
of greatly enhanced pressure which is out of equilibrium
with the local gravitational potential. This results in a rapidly expanding,
roughly ellipsoidal shock wave (Mach $\sim$ 5.5) which heats the
surrounding gas, carrying away from the core the excess thermal energy
resulting from the collision. This shock, which was seen also in the
simulations of Schindler and M\"uller (1993)
(who described it as `lens-shaped'), propagates
more slowly in the region which was preheated by the initial pair of shocks,
leading to a wishbone-shaped distortion in the ellipsoidal shock front.
This is most clearly visible in the plots in the second row of
Figures \ref{Fig:Run A Full XY Slice Plots} and
\ref{Fig:Run A Zoom XY Slice Plots}, which depict the simulation at
$t=4t_{\rm sc}$ and $t=5t_{\rm sc}$.
At this point it has slowed by about a factor of two from its initial speed
relative to the unshocked gas, but because of the temperature gradient its
Mach number has increased to about 18.

In front of the ellipsoidal shock wave the
initial pair of roughly planar shocks has continued to move parallel to the
collision axis. Behind the ellipsoidal shock front, however, these shocks
have mostly been disrupted, leaving behind a disk of relatively cold
gas perpendicular to the collision axis. By $t=5t_{\rm sc}$ this gas has
begun to fall back onto the merger remnant.
The gas outside the region preheated by the initial pair of
shocks (but now inside the
ellipsoidal shock) has been shocked into radial motion away from the
center of the grid, but is decelerating due the gravitational pull of the
(now combined) cluster cores.

By $t=6t_{\rm sc}$ the ellipsoidal shock has left the grid, taking with it
approximately 15\% of the initial mass of the clusters.
The material left in its wake has begun to fall back onto the merger remnant,
creating a weak cylinder-shaped accretion shock.
The density inside the
accretion shock is settling into a roughly spherical distribution, with small
fluctuations in the velocity field. The temperature at the center is slightly
higher than in the regions just inside the accretion shock. As time passes
this temperature difference decreases.
The central density and temperature after $t=7t_{\rm sc}$ are roughly
constant at about twice their initial values.

\begin{figure}
  \figurenum{7}
  \plotone{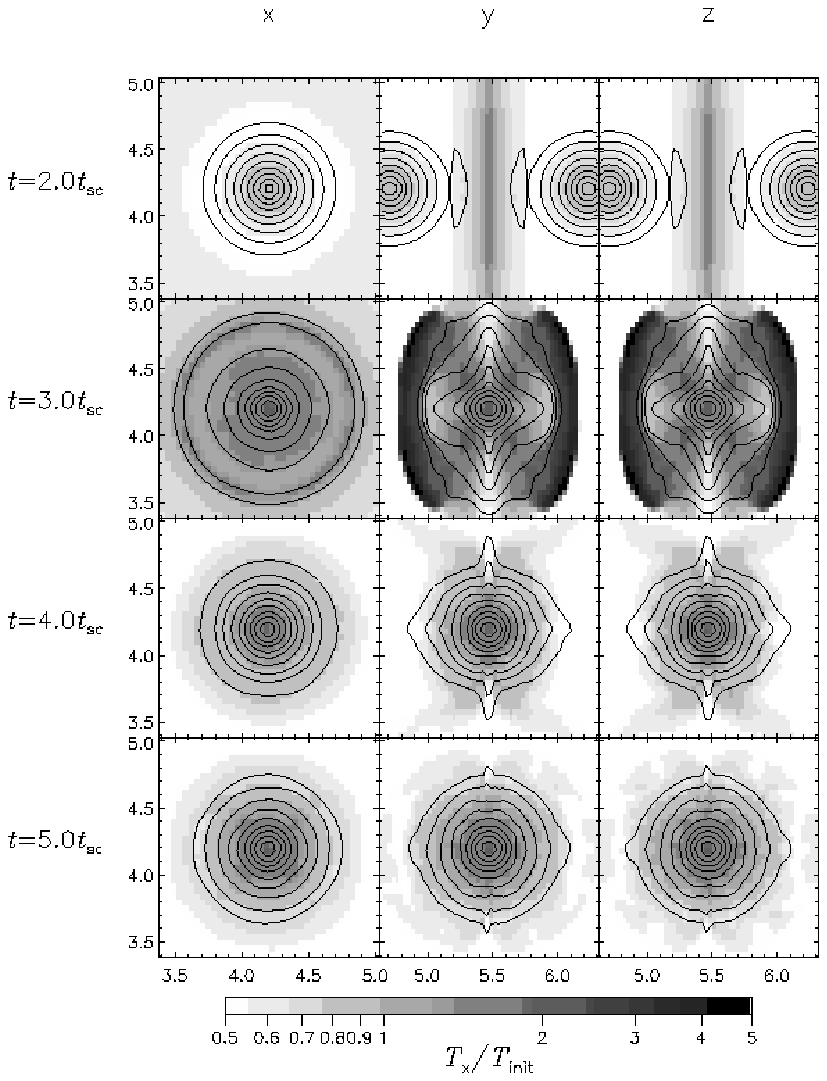}
  \caption{Projected maps for run A of ROSAT X-ray surface brightness ($S_x$)
  and ASCA emission-weighted temperature ($T_x$), viewed along the $x$, $y$,
  and $z$ axes. The innermost $1.6R$ of the grid is shown. $S_x$ (in
  units of $M t_{\rm sc}^{-3}{\rm sr}^{-1}$) is represented by contours spaced
  by a factor of three, with the outermost contour indicating $\log\ S_x =
  -3.5$.
  $T_x$ is represented by shading.}
  \label{Fig:Run A X-ray Projections}
\end{figure}

In Figure \ref{Fig:Run A X-ray Projections} we present projected X-ray
surface brightness contour
maps (integrated over the ROSAT energy band, 0.1--2.4 keV)
for the merging system at four of the representative times of Figure
\ref{Fig:Run A Full XY Slice Plots}, as viewed along the $x$-, $y$-, and
$z$-axis.
For the purpose of generating these maps we have assumed a cluster
mass $M = 4.4\times 10^{14}M_\odot$, radius $R = 1.58$ Mpc,
sound crossing time $t_{\rm sc} = 1.5$ Gyr,
and initial temperature $T_{\rm init} = 4.5\times 10^7$ K.
The bremsstrahlung emissivity $\epsilon_{\rm ff}$
is given by (\cite{RybLig79})
\begin{equation}
\epsilon_{\rm ff} = KI(T)\left({\rho\over MR^{-3}}\right)^2
		         \left({T\over T_{\rm init}}\right)^{1/2}
		         \ MR^{-1}t_{\rm sc}^{-3}\ ,
\label{Eqn:Bremsstrahlung Emissivity}
\end{equation}
\noindent where, for our choices of $G$ and $\mu/k$,
\begin{equation}
K \equiv 0.23 \left({R\over{\rm Mpc}}\right)^{-1}\ .
\end{equation}
\noindent We define
\begin{equation}
I(T) \equiv \int_{T_a/T}^{T_b/T} x^{-0.4} e^{-x}\,dx\ ,
\end{equation}
\noindent where $T_a = 0.1{\rm\ keV}/h$ and $T_b = 2.4{\rm\ keV}/h$
define the observing energy range.
(With the ROSAT energy band, $I(T)$ is roughly constant and
equal to unity for $5\times 10^6{\rm\ K} < T < 10^8{\rm\ K}$.)
As Schindler and M\"uller noted for their simulations, here we see that
a favorable viewing angle is needed to detect the ongoing collision;
we note also that one must observe the system within one sound-crossing
time or so of the core collision to see any significant distortions in
the X-ray isophotes, even for the most favorable viewing angles (along
the $y$ and $z$ axes). During this period the isophotes in the
center of the map have a distinctly oval appearance. However, despite
the strength of the ellipsoidal shock, the density enhancement just behind
it is not sufficient to make it even as luminous as $10^{-5}$ times the
central luminosity. The accretion shock is also not visible.
In the X-ray projections at $t=5t_{\rm sc}$ the cluster appears
spherically symmetric and relaxed from all directions.

Projected temperature maps, weighted by the X-ray emission in the ASCA band
(1.5--11 keV), are shown as grayscales in Figure \ref{Fig:Run A X-ray Projections}.
To generate these maps we assumed the same values for $M$, $R$, $t_{\rm sc}$,
and $T_{\rm init}$ as we used in generating the X-ray surface brightness maps.
The temperature maps make it clear that, in order to see the shocks in this
problem (except for the initial pair), one must look not at the X-ray
surface brightness but at the X-ray temperature; all of the shocks described
earlier are visible in these maps. This is true partly for the following reason.
The bremsstrahlung emissivity (\ref{Eqn:Bremsstrahlung Emissivity})
is very sensitive to the density and only weakly sensitive to the temperature.
However, the Rankine-Hugoniot conditions, which relate
the jumps in density, pressure, and velocity across a shock to the shock's
Mach number $M$ relative to the pre-shock gas, limit the density jump across
the shock to (\cite{LanLif87})
\def\Machlim{\renewcommand{\arraystretch}{0.5}
	     \begin{array}[t]{c}\longrightarrow \\
				\scriptstyle M\to\infty\end{array}
	     \renewcommand{\arraystretch}{1.0}}
\begin{equation}
{\rho_2\over\rho_1} = { (\gamma+1)M^2 \over
                        (\gamma+1) + (\gamma-1)(M^2-1) }
                      \Machlim { \gamma+1 \over \gamma-1 } \,.
\label{Eqn:Density Ratio}
\end{equation}

\noindent Here $\rho_1$ is the
pre-shock density, and $\rho_2$ is the post-shock density.
On the other hand the temperature jump can increase without bound:
\begin{equation}
\begin{array}{l}
\displaystyle
{T_2\over T_1} = { \left[(\gamma+1)+2\gamma(M^2-1)\right] \over
                   (\gamma+1)^2M^2 } \times \\
\\
		\qquad\ \ \left[(\gamma+1)+(\gamma-1)(M^2-1)\right] \\
\\
\displaystyle\quad\Machlim { 2\gamma(\gamma-1) \over (\gamma+1)^2 }M^2
\end{array}
\label{Eqn:Temperature Ratio}
\end{equation}

\noindent Naively substituting the limiting density (\ref{Eqn:Density Ratio})
and temperature (\ref{Eqn:Temperature Ratio}) ratios into
the expression (\ref{Eqn:Bremsstrahlung Emissivity})
for the bremsstrahlung emissivity,
we find that the limiting emissivity enhancement across a strong shock is
\begin{equation}
{ \epsilon_{{\rm ff,}2} \over \epsilon_{{\rm ff,}1} } \Machlim
                  { (\gamma+1)^{2.2}\over(2\gamma)^{0.1}(\gamma-1)^{2.1}}
                  J(T_1) M^{-0.1}\ ,
\label{Eqn:Emissivity Limit}
\end{equation}

\noindent where
\begin{equation}
J(T_1) \equiv {1\over I(T_1)}\int^{T_b/T_1}_{T_a/T_1} dx\, x^{-0.4}
       = {T_b^{0.6}-T_a^{0.6}\over 0.6 T_1^{0.6} I(T_1)}
\end{equation}

\noindent is a weak function of the pre-shock temperature.
Hence X-ray emissivity (even without projection effects) is a much poorer
detector of strong planar shocks than gas temperature. It should be noted,
however, that this result is only valid at large distances from the center
of the cluster potential.


\subsection{$b_{\rm init} = 5r_c$}

\begin{figure}
  \figurenum{8a}
  \plotone{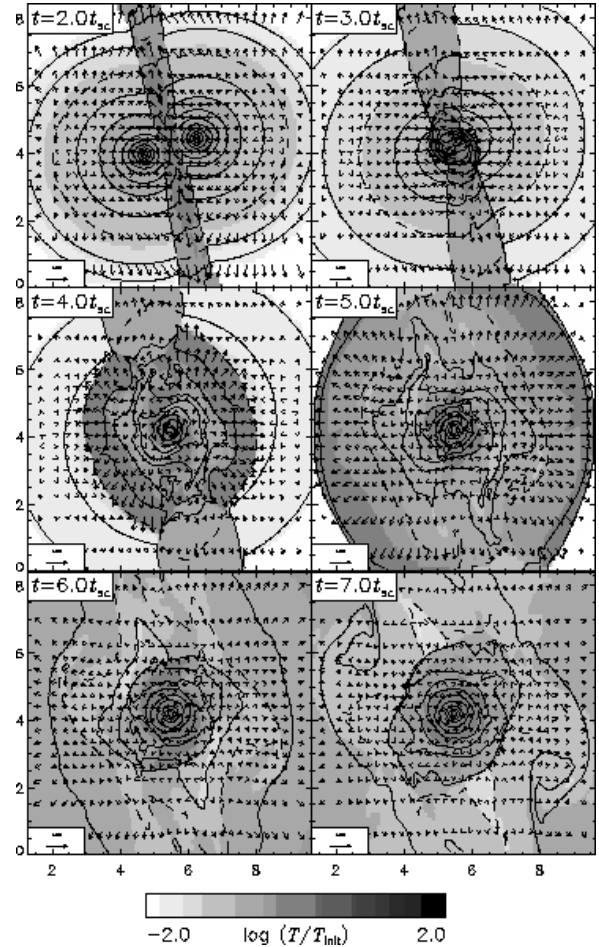}
  \caption{Slices of log density and log
  temperature for run B, taken perpendicular to the $z$-axis through
  the center of the box. Units, ranges, and contour spacing are as in Figure
  \protect\ref{Fig:Run A Full XY Slice Plots}.}
  \label{Fig:Run B Full XY Slice Plots}
\end{figure}

\begin{figure}
  \figurenum{8b}
  \plotone{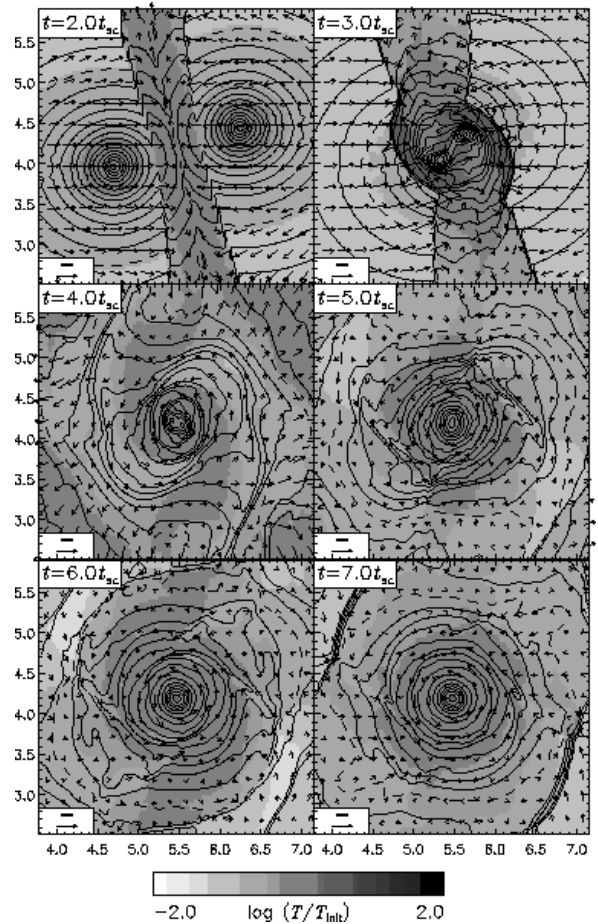}
  \caption{Detail of the innermost
  16\% of Figure \protect\ref{Fig:Run B Full XY Slice Plots}
  (the $xy$-plane). Units, ranges, and contour spacing are as in Figure
  \protect\ref{Fig:Run A Zoom XY Slice Plots}.}
  \label{Fig:Run B Zoom XY Slice Plots}
\end{figure}

\begin{figure}
  \figurenum{8c}
  \plotone{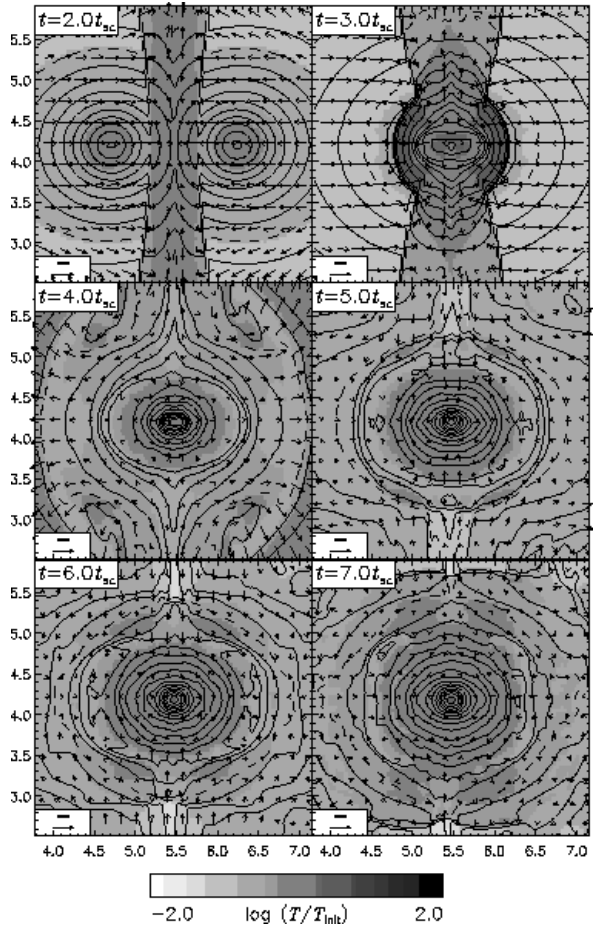}
  \caption{Slices of log density and log
  temperature for run B, taken perpendicular to the $y$-axis through
  the center of the box. Detail of the innermost 16\% of the grid is shown.
  Units, ranges, and contour spacing are as in Figure
  \protect\ref{Fig:Run A Zoom XY Slice Plots}.}
  \label{Fig:Run B Zoom XZ Slice Plots}
\end{figure}

Run B was the first off-center collision, with impact parameter
$b_{\rm init} = 5r_c$. Figures \ref{Fig:Run B Full XY Slice Plots} and
\ref{Fig:Run B Zoom XY Slice Plots} show snapshots of the density,
temperature, and velocity in this run for 100\% and the innermost 16\%,
respectively, of the midplane perpendicular to the $z$-axis (the collision
plane). Once again the abscissa in these plots is the $x$-axis, and the
ordinate is the $y$-axis.
Figure \ref{Fig:Run B Zoom XZ Slice Plots} gives a zoomed view of the
midplane perpendicular to the $y$-axis; here the ordinate is the $z$-axis.

In this run, as in the head-on case, we see a sequence of initial planar
shocks, followed by a fast ellipsoidal
shock as the cores collide, followed by an accretion shock as material
falls back onto the merger remnant. However, the
behavior of these shocks is more complex, and the
presence of angular momentum in the system alters their roles somewhat.
The cluster cores survive intact longer, and the merger remnant requires more
time to reach equilibrium.

The planar shocks produced by the clusters' initial interaction
(seen at $t=2t_{\rm sc}$) are now oblique, making a $\sim 72^\circ$ angle with
the collision axis. The temperature jump and outflow velocity in the
post-shock region are comparable to the values in run A.

As previously noted, in the off-center runs the cluster cores
survive intact beyond the time at which they would have merged in the
head-on case. The size of each core is determined roughly by the
ram pressure of the gas through which it passes: the cluster remains
intact where its own gas pressure exceeds the ram pressure. As we increase
the impact parameter of the collision, the density $\rho'$ of the gas
through which each core must pass decreases rapidly, while the relative
speed $v$ remains roughly the same. The core's gas pressure $p$ decreases
with radius in the same way that $\rho'$ decreases with $b_{\rm init}$.
Therefore the radius at which $p$ drops below the ram pressure
($\sim \rho' v^2$) should be roughly proportional to $b_{\rm init}$.
The sizes of the cluster cores seen in runs B and C are consistent with
this explanation.

Because the planar shocks are now oblique they play a new dynamical role
not found in the head-on collision.
The cluster cores, defined as described above, carry most of the
angular momentum and survive past the point at which they would have
coalesced in the head-on case. The cores are not disrupted by the planar
shocks; instead each passes through the shock which propagated toward it,
then drives the other shock ahead of it, so that the central part of each
planar shock is twisted into a spiral pattern. As each core moves through
the outer regions of the other cluster, it creates a strong,
curved bow shock which propagates out of the center of the grid. Together
these form a fast ellipsoidal shock front similar to (but weaker than)
that created by the core collision in the head-on case.
This shock serves to dissipate the
excess kinetic energy of the cluster cores, enabling them to fall into orbit
about one another. However, not all of the excess is carried away by this
shock; the remainder is dissipated more slowly by what is left of the
slower, originally planar shocks ($t=3t_{\rm sc}$).
By turning the material passing through them,
these spiral shocks gradually transfer the orbital
angular momentum of each core to the surrounding low-density gas. Shortly
after $t=4t_{\rm sc}$ the cores lose enough angular momentum in this way to fall
almost directly in toward each other, having completed nearly a full revolution.

By $t=5t_{\rm sc}$ the ellipsoidal shock has nearly left the grid, and the gas
in its wake has begun to fall back onto the merger remnant, forming an accretion
shock. By $t=7t_{\rm sc}$ this accretion shock is fairly well-developed and
lies roughly $1.5R$ from the center of the remnant.
The remnant itself is rotating, and weak spiral shocks (the remains
of the twisted planar shocks described earlier) can be seen; these are
completing the redistribution of angular momentum out of the cluster cores.
By $t=7t_{\rm sc}$ the merger remnant has relaxed to a roughly ellipsoidal shape
and is rotating about the $z$-axis. A small amount of density substructure
is still present in the innermost part of the remnant ($r<r_c$) at this time,
although on this scale the core of the remnant appears to be isothermal.

\begin{figure}[tbh]
  \figurenum{9}
  \plotone{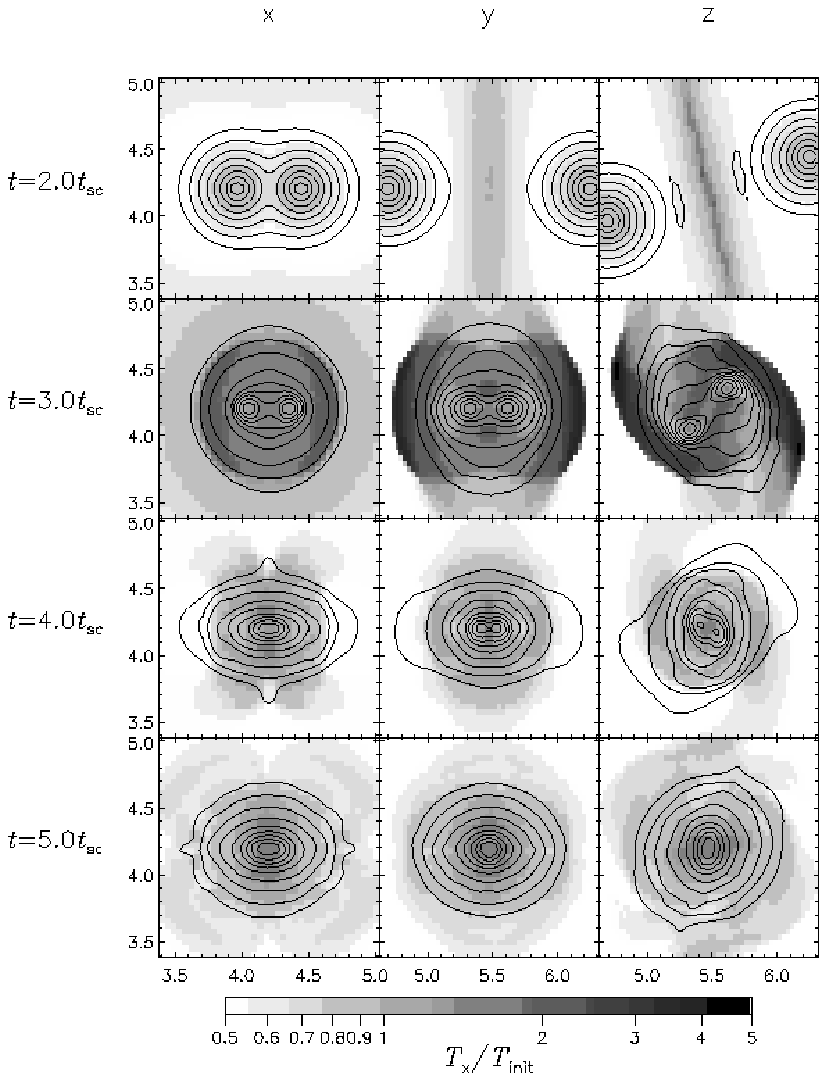}
  \caption{Projected maps for run B of ROSAT X-ray surface brightness ($S_x$)
  and ASCA emission-weighted temperature ($T_x$), viewed along the $x$, $y$,
  and $z$ axes. The innermost $1.6R$ of the grid is shown. $S_x$ (in
  units of $M t_{\rm sc}^{-3}{\rm sr}^{-1}$) is represented by contours spaced
  by a factor of three, with the outermost contour indicating $\log\ S_x =
  -3.5$.
  $T_x$ is represented by shading.}
  \label{Fig:Run B X-ray Projections}
\end{figure}

The projected ROSAT X-ray surface brightness for run B (contours in Figure
\ref{Fig:Run B X-ray Projections}) shows a clearly separated double-peak
structure while the cluster cores are orbiting one another.
The bending of the planar shocks is also visible. After the cores collide
at $t\sim 4t_{\rm sc}$, the bimodal structure disappears, but
the X-ray isophotes continue to appear significantly distorted,
forming a barlike structure which rotates about the $z$-axis. When viewed
along the bar, the cluster appears roughly spherical, and when viewed
off-axis in the plane of the bar's rotation, the cluster appears elliptical,
with the major axes of the isophotes roughly parallel. It is only when the
cluster is viewed from well outside the plane of the bar's rotation that the
differing ellipticities and angles of orientation of the isophotes reveal
the ongoing equilibration of the cluster. However, this cluster continues to
appear aspherical from some directions long after the cluster formed in the
head-on case has ceased to show any detectable evidence of a collision.

The projected ASCA-weighted temperature for this run (grayscale in Figure
\ref{Fig:Run B X-ray Projections}) shows much clearer evidence for an
off-center collision, but after the cores have collided this also requires
a favorable viewing angle. As in the head-on case, a hot bar-shaped structure
is present in the temperature maps. However, here it is perpendicular to the
line connecting the cluster cores, not to the initial collision axis.


\subsection{$b_{\rm init} = 10r_c$}

\begin{figure}[tbh]
  \figurenum{10a}
  \plotone{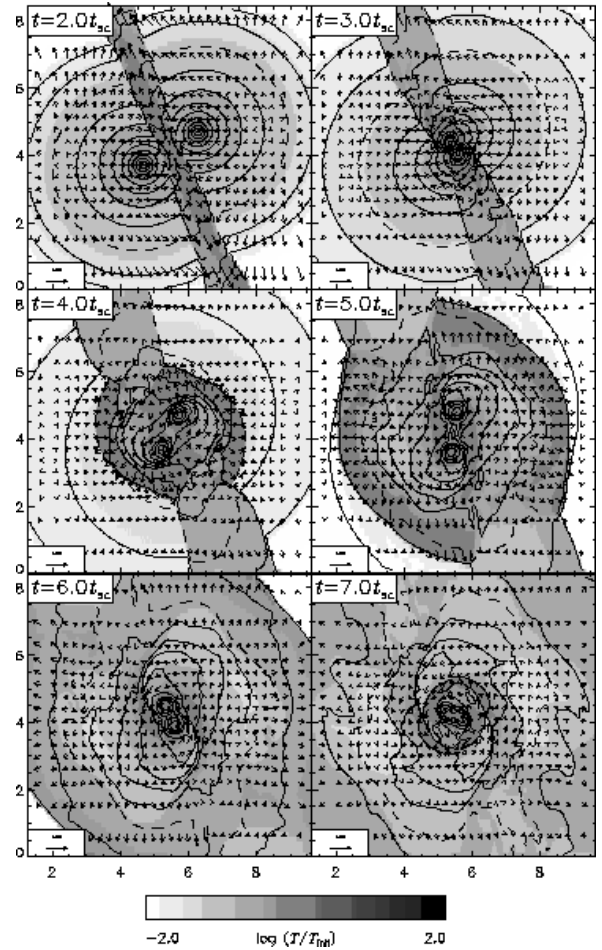}
  \caption{Slices of log density and log
  temperature for run C, taken perpendicular to the $z$-axis through
  the center of the box. Units, ranges, and contour spacing are as in Figure
  \protect\ref{Fig:Run A Full XY Slice Plots}.}
  \label{Fig:Run C Full XY Slice Plots}
\end{figure}

\begin{figure}
  \figurenum{10b}
  \plotone{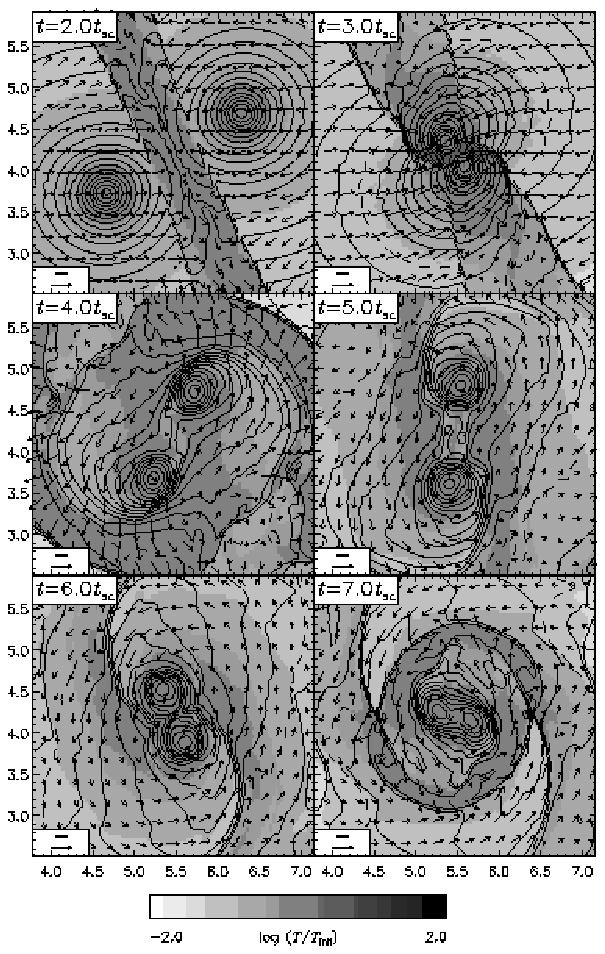}
  \caption{Detail of the innermost
  16\% of Figure \protect\ref{Fig:Run C Full XY Slice Plots}
  (the $xy$-plane). Units, ranges, and contour spacing are as in Figure
  \protect\ref{Fig:Run A Zoom XY Slice Plots}.}
  \bigskip\bigskip
  \label{Fig:Run C Zoom XY Slice Plots}
\end{figure}

\begin{figure}
  \figurenum{10c}
  \plotone{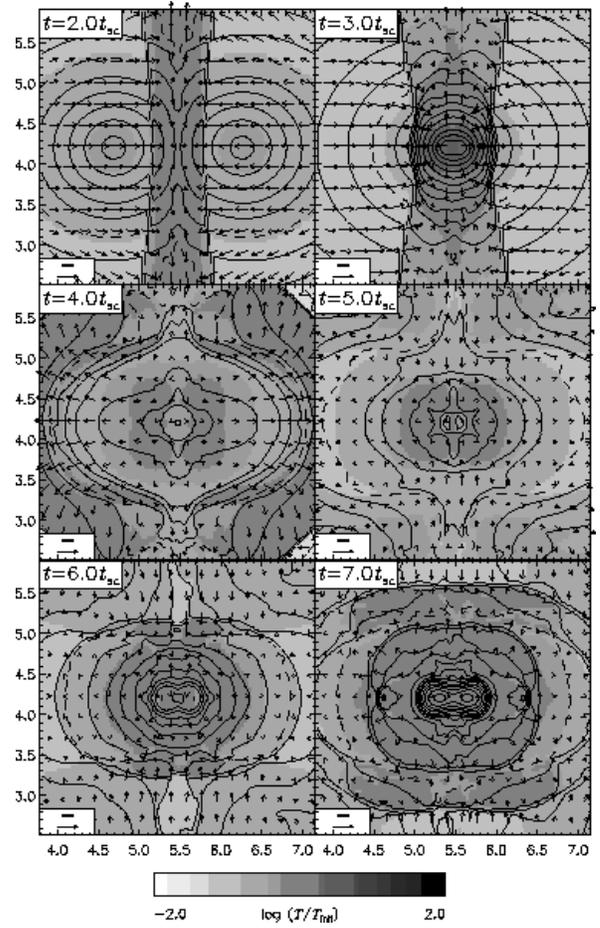}
  \caption{Slices of log density and log
  temperature for run C, taken perpendicular to the $y$-axis through
  the center of the box. Detail of the innermost 16\% of the grid is shown.
  Units, ranges, and contour spacing are as in Figure
  \protect\ref{Fig:Run A Zoom XY Slice Plots}.}
  \label{Fig:Run C Zoom XZ Slice Plots}
\end{figure}

In Run C we used an impact parameter $b_{\rm init} = 10r_c$. For this run,
Figures \ref{Fig:Run C Full XY Slice Plots} and
\ref{Fig:Run C Zoom XY Slice Plots} show full and zoomed snapshots
of the midplane perpendicular to the $z$-axis (the collision
plane). Figure \ref{Fig:Run C Zoom XZ Slice Plots} gives a zoomed view of the
midplane perpendicular to the $y$-axis.
This run displays many of the same qualitative features as the $b_{\rm init}
= 5r_c$ case. However, while the collision itself is more violent, and
the merger remnant requires much longer to relax to a new equilibrium
configuration, the release of kinetic energy is much more gradual.

The initial interaction of the clusters' outer regions again forms a pair of
planar shocks, now at a $\sim 54^\circ$ angle with the collision axis.
Because of their greater obliqueness they do not begin to twist until
much later than in run B, and they brake the cluster cores and transfer
their angular momentum to the surrounding gas less efficiently.
The ellipsoidal shock which allows the cores to remain bound to each other
does not
form until just after $t=3t_{\rm sc}$. In this run it dissipates a smaller
fraction
of the cores' initial energy than in the other two cases. A larger fraction
of this energy is dissipated by the spiral shocks which are driven by the
cores as they orbit each other.
The cores complete about 1 1/2 orbits in the process of spiralling in
toward each other; they do not merge until $t=6.5t_{\rm sc}$.
The size of the cores themselves is
much larger (diameter $\sim 8r_c$ vs.\ $\sim 5r_c$ for $b_{\rm init}=5r_c$),
supporting the idea that their size is determined by ram pressure.

\begin{figure}
  \figurenum{11}
  \plotone{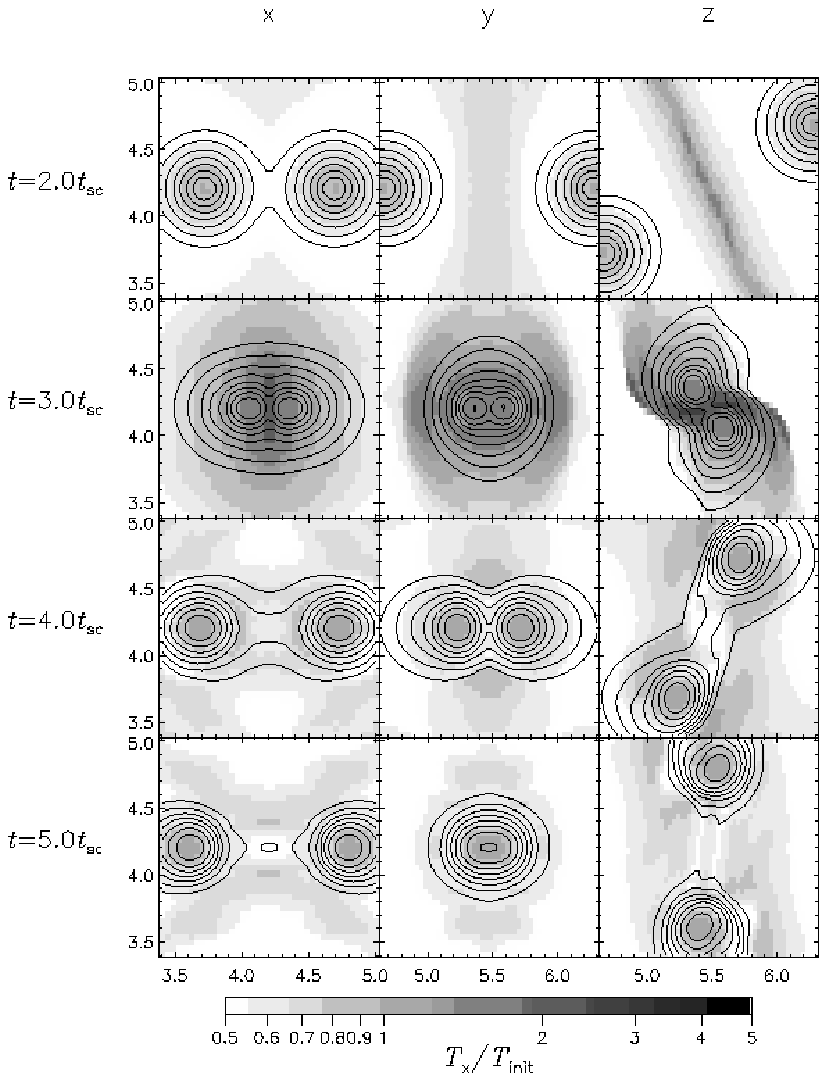}
  \caption{Projected maps for run C of ROSAT X-ray surface brightness ($S_x$)
  and ASCA emission-weighted temperature ($T_x$), viewed along the $x$, $y$,
  and $z$ axes. The innermost $1.6R$ of the grid is shown. $S_x$ (in
  units of $M t_{\rm sc}^{-3}{\rm sr}^{-1}$) is represented by contours spaced
  by a factor of three, with the outermost contour indicating $\log\ S_x =
  -3.5$.
  $T_x$ is represented by shading.}
  \label{Fig:Run C X-ray Projections}
\end{figure}

The projected ROSAT X-ray emission for this run (contours in Figure
\ref{Fig:Run C X-ray Projections}) shows significantly more evidence for
an ongoing merger than in the other two runs. Bimodal structure is present in
the X-ray maps well past $t=5t_{\rm sc}$. In the projected
ASCA emission-weighted temperature maps (grayscale in Figure
\ref{Fig:Run C X-ray Projections}), bimodal structure also persists past
$t=5t_{\rm sc}$. The bow shocks are easily visible at $t=4t_{\rm sc}$ and
$5t_{\rm sc}$.
From above the plane of the bar's rotation, some evidence of spiral structure
can be seen at $5t_{\rm sc}$.

\section{Discussion}

In this section we compare the three collision runs described in the
previous section in terms of the time required for each to relax to
its final equilibrium state, the increase in X-ray luminosity during
the collision, and the structure of the merger remnant.


\subsection{Equilibration time}

In hierarchical models of large-scale structure formation, clusters of
galaxies form through a series of mergers followed by periods of relaxation.
How relaxed clusters look in such a model at any given time depends on
the ratio of the equilibration time to the average interval between mergers.
If this ratio is much larger than one, so that, on average, clusters have not
completely
assimilated the effects of one merger before the next takes place, clusters
will not appear to be in equilibrium. If the opposite is true, so that the
equilibration time is much smaller than the merger interval, then clusters
will usually appear to be virialized, even if the merger rate is quite high.
The fraction of clusters which appear to be far from equilibrium therefore is
telling us something about this ratio.

The rate at which mergers take place depends on the
assumed underlying cosmological model, while
the average equilibration time consists of two factors: the equilibration time
for mergers having particular values of the impact parameter, relative
velocity, and relative mass, and the distribution of mergers with these
parameters as a function of cosmological parameters. The first factor is
presumably only weakly dependent on cosmology, since the relaxing region
has detached itself from the general cosmic expansion; the second depends
on the tidal torques generated in different models. If we are to reason
backward from the fraction of clusters which appear unrelaxed and the
rate at which clusters are observed to merge to deduce the underlying
cosmological model (as specified by $\Omega_0$, $\Lambda$, etc.), we need
to understand the physical factors influencing the equilibration rate for
collisions with given parameters.

While our present simulations concern themselves only with the behavior
of the gaseous component of galaxy clusters, thus ignoring the effect on
the equilibration rate of energy transfer between the gas and the dark matter,
we can nevertheless draw some conclusions regarding the rate of equilibration
after an off-center merger.

Physically motivated definitions for the equilibration time will
differ markedly from those based on observational criteria;
furthermore, physically motivated definitions based on average quantities
will differ from those based on the actual presence of substructure
(regardless of whether this is observable or not). In general we find that
observational criteria (such as the projected luminosity and temperature
maps described in the previous section) lead to shorter equilibration
times than criteria based on the degree of virialization of the
merger remnant.

From an observational standpoint the time required in our simulations for the
merging clusters to reach a new equilibrium state is short. The projected X-ray
surface brightness (Figures \ref{Fig:Run A X-ray Projections},
\ref{Fig:Run B X-ray Projections}, and \ref{Fig:Run C X-ray Projections})
shows little evidence of the shocks which are produced in each run.
The high density of material in the cluster cores' gravitational wells
accounts for most of the X-ray emission, 
and what evidence exists in the X-ray maps for an ongoing merger,
such as bimodality, is only present when the cores are well-separated
in the plane of the sky, or when they have not yet fully coalesced.
This is partly due to projection effects and partly because the
bremsstrahlung emissivity poorly delineates planar shocks.

In the head-on case bimodality is only present before the collision;
after the collision the anisotropy caused by material falling back onto
the remnant perpendicular to the collision axis is only visible for about
one sound crossing time.
In the case of $b_{\rm init} = 5r_c$ the cores do not coalesce as rapidly;
consequently they are distinguishable both before the collision and for more
than one sound crossing time after the ellipsoidal shock front forms.
During this time they are orbiting one another, and they only appear
well-separated if viewed from above the orbital plane or if, by chance,
they are viewed at a point in their orbit when neither lies in front of
the other. However, from all directions the X-ray isophotes remain fairly
elliptical,
with an axial ratio of approximately 2:1, until several sound crossing times
have passed. When viewed in the collision plane the long axes of the
isophotes are all oriented in the same direction, parallel to the collision
plane. When viewed from above the collision plane, their axes are not
aligned; the innermost contours are advanced in the direction of rotation
relative to the outermost ones. Initially the difference in angle is as
large as $90^\circ$, but this decreases with time. By $t=5t_{\rm sc}$
the difference has decreased to $45^\circ$, and at late times the remnant
appears spherically symmetric.
Thus the rotation of the merger remnant in run B does not appear
to significantly affect the shape of the X-ray isophotes at late times.
The case of $b_{\rm init} = 10r_c$ displays many of the same characteristics
as that of $b_{\rm init} = 5r_c$ prior to the final merging of the cluster
cores, except that the cores' inspiral requires more than three crossing times,
allowing bimodal structure to be visible substantially longer than in runs A
and B.

\begin{figure}[tbh]
  \figurenum{12}
  \plotone{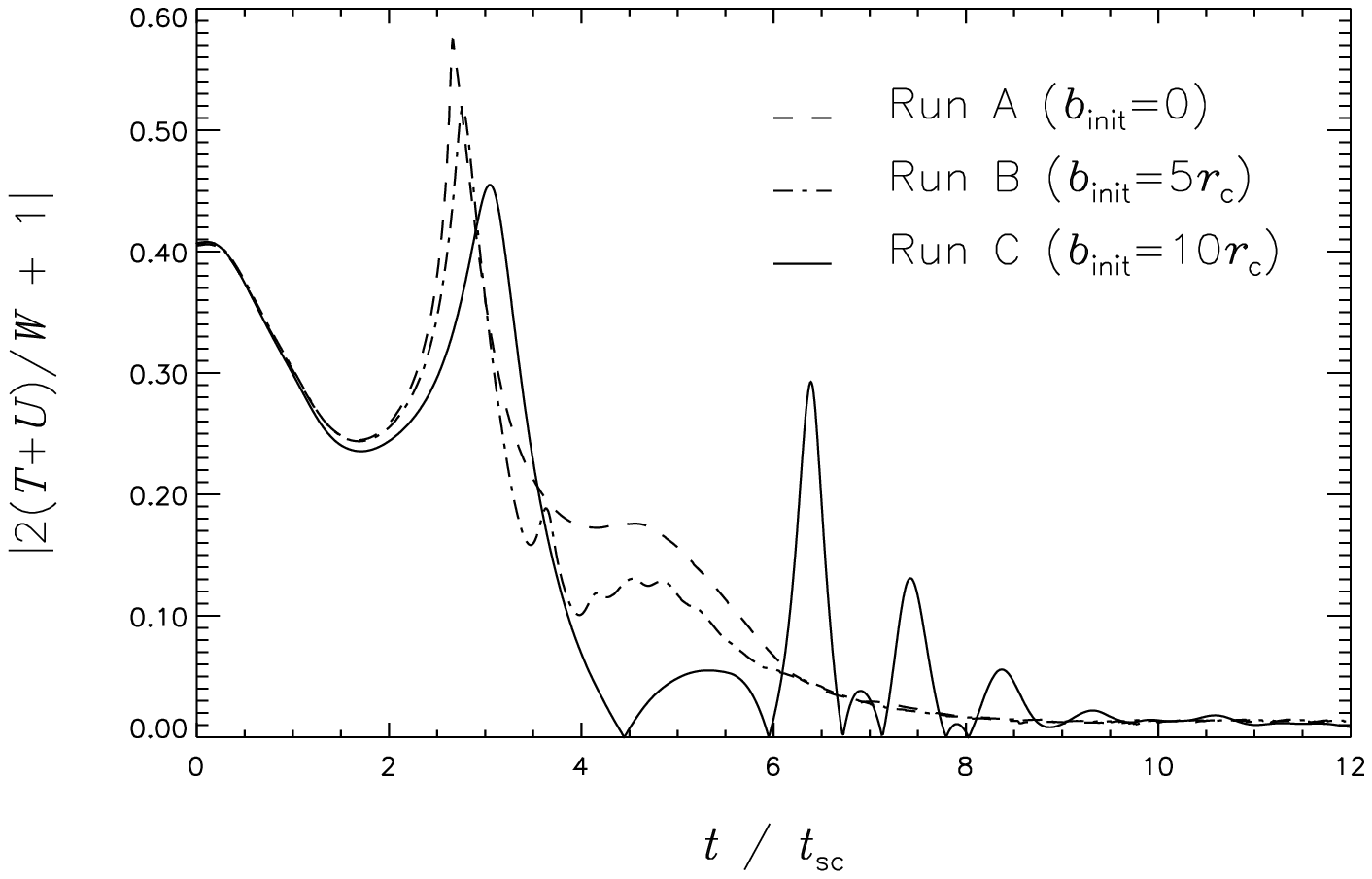}
  \caption{Factor by which the colliding clusters depart from virial equilibrium
  as a function of time for the three collision runs.}
  \label{Fig:Equilibrium Criterion Comparison Plot}
\end{figure}

The virial theorem provides a useful physical criterion for determining when
the system is close to equilibrium. Whether
the system is relaxed or not, the virial theorem requires that
\begin{equation}
{1\over 2}{d^2I\over dt^2} = 2T + W + 3(\gamma-1)U
\label{Eqn:Virial Theorem}
\end{equation}
\noindent (\cite{Cha61}), where $I$ is the moment of inertia and $T$, $W$,
and $U$ are the total kinetic, potential, and thermal energies,
respectively:
\begin{eqnarray}
\nonumber
I & \equiv & {1\over 2}\int d^3x\,\rho({\bf x}) |{\bf x}|^2 \\
\nonumber
T & \equiv & {1\over 2}\int d^3x\,\rho({\bf x}) |{\bf v}({\bf x})|^2 \\
W & \equiv & -{1\over 2}G\int\int d^3xd^3x'\,{\rho({\bf x})\rho({\bf x'})
		\over |{\bf x} - {\bf x'}|} \\
\nonumber
U & \equiv & {1\over\gamma-1}\int d^3x\, p({\bf x})\ .
\end{eqnarray}
\noindent The volume integrals in these definitions are taken over the entire
grid. When the system has reached a steady state, the left-hand side
of equation (\ref{Eqn:Virial Theorem}) will be zero. We can approximate this
criterion for practical purposes by dividing equation (\ref{Eqn:Virial
Theorem}) by $W$:
\begin{equation}
\left|{ {1\over 2W}{d^2I\over dt^2} }\right| =
\left|{ 2{T\over W} + 1 + {3(\gamma-1)U\over W} }\right| < \epsilon \ll 1\ .
\label{Eqn:Equilibrium Criterion}
\end{equation}
\noindent When this criterion is satisfied for some suitably chosen value of
$\epsilon$, the system will be close to equilibrium. In Figure
\ref{Fig:Equilibrium Criterion Comparison Plot} we have plotted the
second expression in equation (\ref{Eqn:Equilibrium Criterion}) as a function
of time for the three runs. We set $\epsilon$ to $0.02$,
roughly the level of round-off error in our determination of $T$, $U$, and
$W$, and roughly the point at which the virial parameter becomes constant in
these runs. Although during the period following the collision
the behavior of the energy in run A differs
substantially from that in run B, both runs reach equilibrium
according to our criterion approximately $5t_{\rm sc}$ after the initial core
interaction (represented by the initial peak for each curve). Run C, in
contrast, requires more than $6t_{\rm sc}$ to reach equilibrium after the
initial core interaction. The virial parameter in this case
passes through several maxima, reflecting the strength of the second
and subsequent core interactions.

Using observational criteria to determine equilibration
time, we see that off-center collisions can take substantially longer
than head-on collisions to reach a new equilibrium state.
The virial equilibration time is less sensitive to the impact parameter
but is at least twice the observational equilibration time. Given typical
sound crossing times $t_{\rm sc} \sim 1{\rm\ Gyr}$, both the observational
and the virial equilibration times can be
a substantial fraction of the age of the universe.

Adding collisionless dark matter to our simulations would
tend to increase the equilibration time, since the dark matter does not
experience shocks and so must dissipate its initial kinetic energy through
violent relaxation and drag due to the baryonic contribution to the
potential. Adding radiative cooling would tend to decrease the equilibration
time somewhat by permitting the gas in the cluster cores to radiate away
some of the kinetic energy in excess of the potential energy available to
bind the cores, leading to weaker shocks during the collision. However,
the cooling time is typically very long outside the cluster cores, so
cooling should not affect the dynamics of these regions very much.
In the central region where the cooling time is relatively short,
the timescale for production of the ellipsoidal shock, which
dissipates most of the cores' energy, is also short. Therefore it is
likely that radiative cooling will have a much smaller effect on the
equilibration timescale than it will have, for instance, on the luminosity
of the clusters during the collision or the density and temperature profiles
of the merger remnant.


\subsection{Brightening during the collision}

\begin{figure}[tbh]
  \figurenum{13}
  \plotone{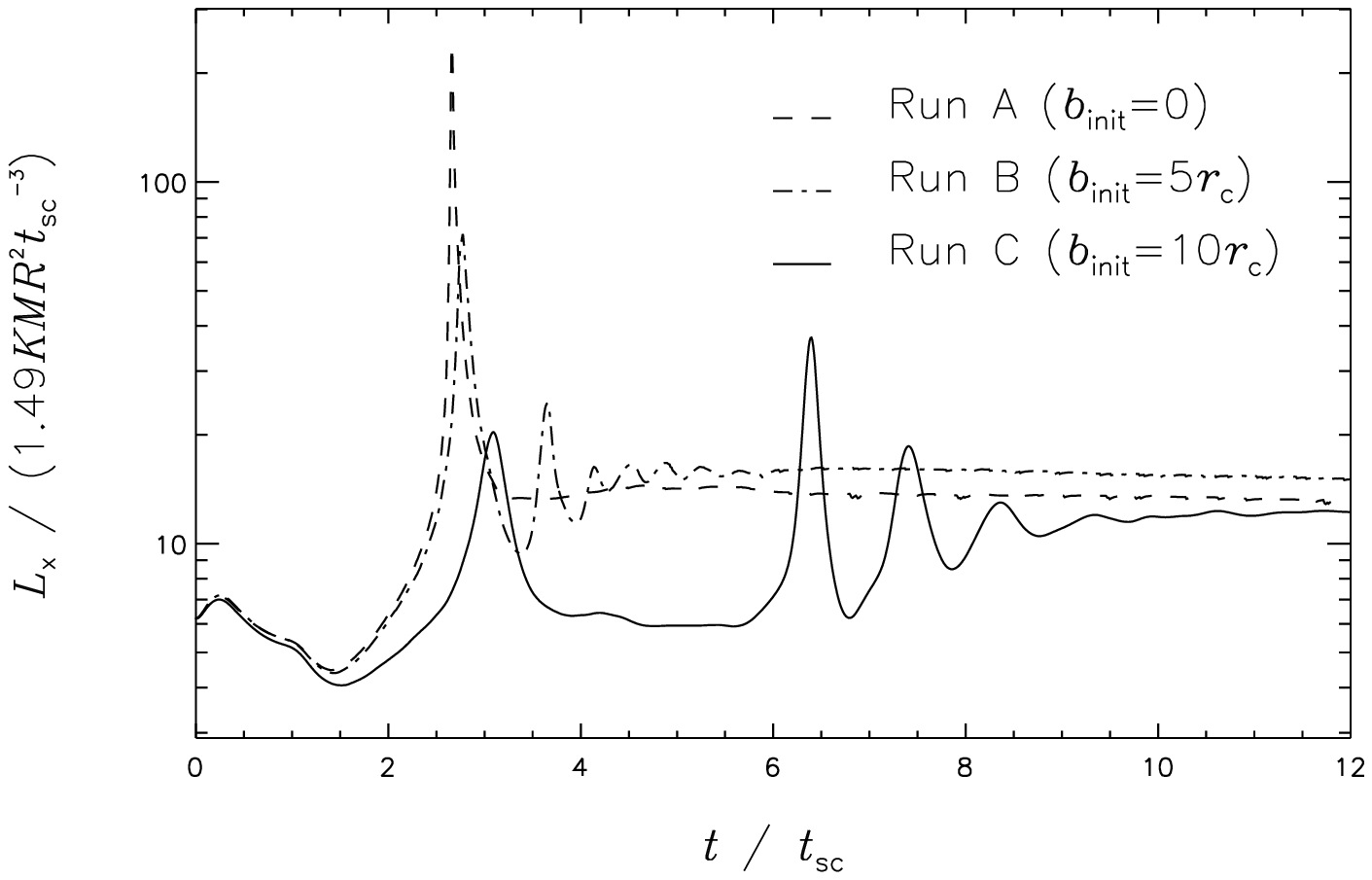}
  \caption{Radiative energy loss rates due to bremsstrahlung emission for the
  three collision runs (were cooling included).}
  \label{Fig:Comparative Luminosity Plot}
\end{figure}

During the course of each collision the X-ray luminosity
of the system varies significantly. The effect of changing the impact parameter
on this luminosity evolution can be seen in
Figure \ref{Fig:Comparative Luminosity Plot},
which shows as a function of time the total energy loss rate, integrated over
all frequencies, due to bremsstrahlung emission in each of the collision runs.
This is just the total X-ray luminosity as a function of time:
\begin{equation}
L_{\rm x} = 1.49 K \sum_{ijk} {\Delta V_{ijk}\over R^3}
	    \left({T_{ijk}\over T_{\rm init}}\right)^{1/2}
	    \left({\rho_{ijk}\over MR^{-3}}\right)^2\ \ MR^2t_{\rm sc}^{-3}\ .
\end{equation}
\noindent Before $t=1.5t_{\rm sc}$ the luminosity drops by about
one-third due to the
adiabatic cooling related to our initial conditions, as discussed in
Section 3. Near $t=3t_{\rm sc}$, when the ellipsoidal shock forms in each run,
the radiative energy loss rate increases
briefly by as much as a factor of 50, then drops to a new constant
level 4--5 times its lowest pre-collision value. The height and width of the
luminosity peak indicate the amount and duration of the compression experienced
by the intracluster gas as the cluster cores approach one another. Thus in
the head-on case we see one tall peak lasting about $t_{\rm sc}/2$, while in
run B the initial peak is about one-third as tall and is followed by several
smaller peaks as the cluster cores spiral inward toward one another. In run C
the cores orbit several times before coalescing; hence we see several
peaks, with the luminosity following each peak gradually increasing to roughly
the same final level as in runs A and B. Because of the long duration of the
first orbit in run C, the second and subsequent luminosity peaks in this case
appear about $3t_{\rm sc}$ after the first peak.

Because we do not include radiative cooling in our simulations, our luminosity
enhancements are much greater than, for example, in the simulations by
Schindler and M\"uller (1993). Nevertheless the results in Figure
\ref{Fig:Comparative Luminosity Plot} provide information about the strength
of the shocks produced in each run which should carry over to simulations
which do include cooling. Our observations suggest that magnitude-limited
surveys of galaxy clusters will be subject to a selection effect based on
merging activity, and in particular to one based on impact parameter.
Head-on mergers produce stronger shocks and are
more luminous than off-center mergers. Mergers are much more
visible for a relatively short time while the cluster cores are first
interacting than either before or after this time. Because
mergers in progress can be much more luminous than quiescent clusters, we
may be overestimating the amount of merging activity at high redshift.
Alternatively, since projection effects can hide isophotal evidence for
mergers at most orientations, we may observe merging systems as highly
luminous, apparently relaxed clusters.


\subsection{Structure of the merger remnant}

For all three runs we were able to follow the
progress of the collision until well after the merger remnant became
virialized. In this section we compare the angle-averaged density and
temperature profiles in each run at $t=12t_{\rm sc}$. We also
examine the rotation of the remnant in runs B and C at $t=12t_{\rm sc}$.

\begin{figure}
  \figurenum{14}
  \plotone{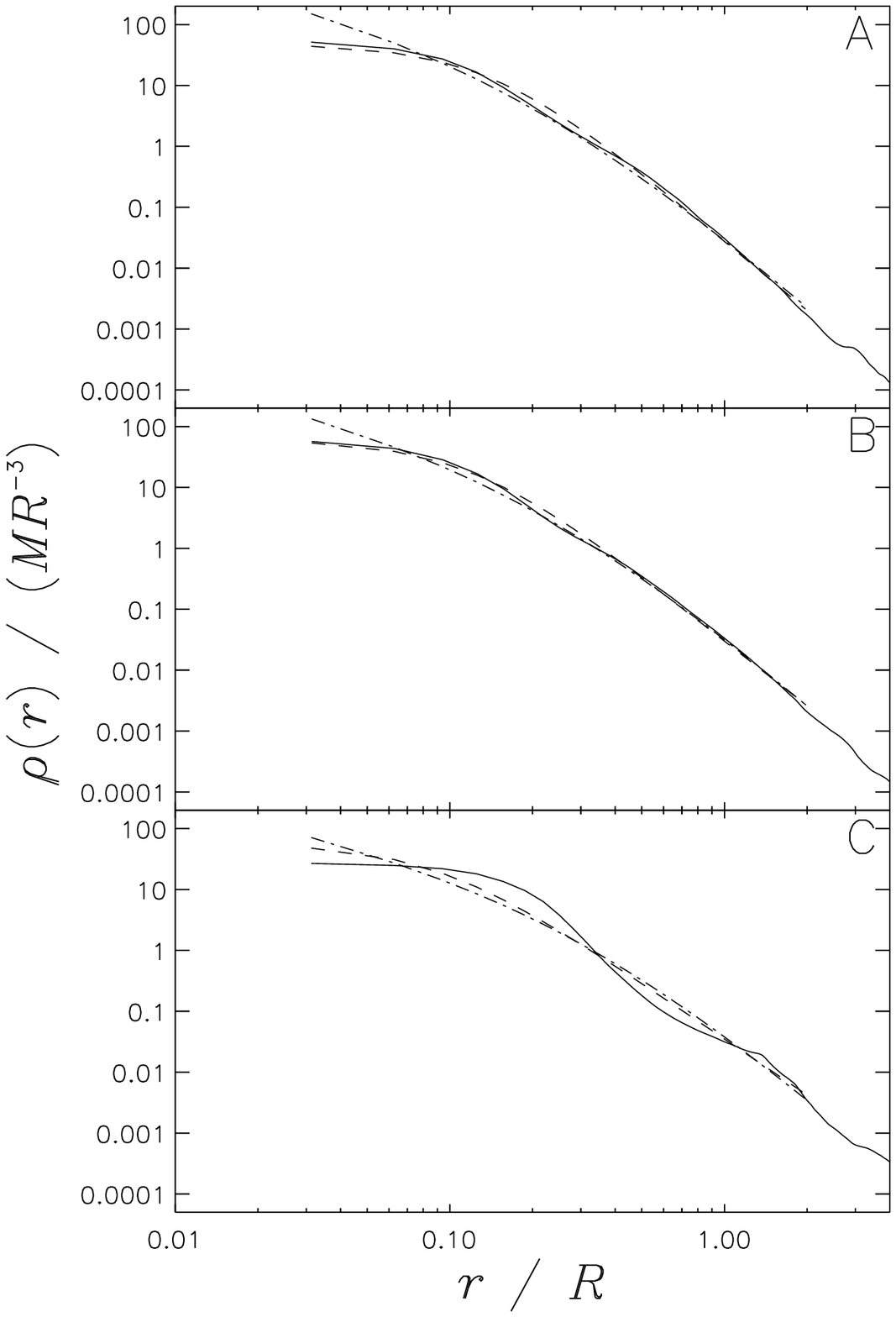}
  \caption{Angle-averaged density profiles, taken about the center of mass, for
  the merger remnants in runs A, B, and C at $t=12t_{\rm sc}$.
  The average profiles
  are plotted as solid lines. Best-fit model profiles described in the text
  include
  the $\beta$-model (dashed lines) and the Hernquist model (dot-dashed lines).}
  \label{Fig:Comparative Density Profile Plot}
\end{figure}

In Figure \ref{Fig:Comparative Density Profile Plot} we plot using solid
lines the angle-averaged
density profiles for runs A, B, and C at $t=12t_{\rm sc}$.
To determine the effect of core heating on the structure of the merger
remnant, we have fitted the average profiles for $r < 2R$
using two models: the $\beta$-model (\cite{CavFF76}),
\begin{equation}
\rho_\beta(r) = {\rho_0 \over \left[1 + (r/r_c)^2\right]^{3\beta/2}}\ ,
\label{Eqn:Beta Model}
\end{equation}
\noindent and the analytical approximation to the de Vaucouleurs (1948) model for
elliptical galaxies described by Hernquist (1990),
\begin{equation}
\rho_H(r) = {M_H\over 2\pi}{a_H\over r}{1\over(r+a_H)^3}\ .
\label{Eqn:Hernquist Model}
\end{equation}
\noindent In Figure \ref{Fig:Comparative Density Profile Plot}
the best-fit $\beta$-model profiles are shown as dashed lines, and the best-fit
Hernquist profiles are shown as dot-dashed lines.
The parameters corresponding to these fits appear in Table
\ref{Table:Density Model Fits}. Note that the initial density profile for each
cluster (equation \ref{Eqn:HubbleLaw}) corresponds to a $\beta$-model
profile with $\beta=1$.

\begin{table*}
\tablenum{2}
\caption{
  Best-fit $\beta$-model and Hernquist model
  parameters for collision runs at $t=12t_{\rm sc}$.
  The first line gives $\beta$-model parameter values for each cluster at $t=0$.
  Note that the best-fit values for run C do not provide a satisfactory fit.
  \label{Table:Density Model Fits}
  }
\begin{center}
\begin{tabular}{lrrrrr}
		&\multicolumn{3}{c}{$\beta$-model}	
		&\multicolumn{2}{c}{Hernquist model}\\
		&\multicolumn{3}{c}{\hrulefill}		
		&\multicolumn{2}{c}{\hrulefill}\\
Run		&$\rho_0/MR^{-3}$	&$r_c/R$	&$\beta$	&$M_H/M$
&$a_H/R$\\
\tableline
original	& 30.5			& 0.111	& 1.00	& ---	& ---\\
A		& 48.2			& 0.142	& 1.27	& 1.60	& 0.184\\
B		& 60.6			& 0.119	& 1.19	& 1.69	& 0.205\\
C		& 56.9			& 0.0904	& 1.02	& 1.74	&
0.304\\
\tableline
\end{tabular}
\end{center}
\end{table*}

Strictly speaking, the $\beta$-model is intended to describe the distribution
of gas which is in equilibrium with a potential determined by a collisionless
matter component, such as galaxies or dark matter; the value of $\beta$
then gives
the ratio of the velocity dispersion of the collisionless component to the
temperature of the gas. Here we have no collisionless component, so $\beta$
serves more as an indicator of deviations from an isothermal distribution
and as a measure of differences in core radius and central density
among the runs. In particular, we expect shock heating during a collision
to lead to a merger remnant with a core radius larger than that of the
original clusters.

The Hernquist model, in contrast to the $\beta$-model,
diverges as $r\rightarrow 0$, and at large $r$ it drops as $r^{-4}$, whereas
fitting with the $\beta$-model allows the slope at large $r$ to vary.
Because of our grid resolution and initial cluster profiles we do not expect
the Hernquist model to give a good fit near $r=0$, but fitting with
it does allow us to
check the asymptotic density slope independently of the shape of the density
profile at small $r$. The asymptotic slope is of interest because $N$-body
simulations of halo formation in cold dark matter (CDM) universes (proceeding,
as it does in such models, through mergers) have yielded differing values
for the slope. Methods neglecting infall from large scales, such as that used
by Dubinski and Carlberg (1991) in studying halo formation in CDM and that used by
Hernquist (1992, 1993) in simulations of galaxy mergers, produce remnants
with asymptotic profiles proportional to $r^{-4}$. Others (e.~g.\ \cite{NFW96})
have argued for an $r^{-3}$ dependence on the basis of simulations which
do include infall. Our calculations differ from these in that we only include
collisional matter and we impose an initial profile with $r^{-3}$ asymptotic
behavior. However, like Dubinski and Carlberg we implement a vacuum
boundary, neglecting infall. If hydrodynamical effects are weak at large radius,
we may therefore be able to check their results. In our single-cluster tests
(Section 3) we find that the asymptotic slope resulting from our artificial
density cutoff at $r=R$ (which would not be needed if we could include infall)
is $r^{-4.8}$, much steeper than $r^{-3}$ or $r^{-4}$, whereas the slope
inside the cutoff radius remains close to $r^{-3}$. We may therefore expect
that any change to an $r^{-4}$ profile in our simulations is an effect of the
collision and not of the initial conditions.

In examining Figure \ref{Fig:Comparative Density Profile Plot} and
Table \ref{Table:Density Model Fits}, we find that in none of the
runs is the merger remnant isothermal; instead, the density profile
in each case drops off more steeply than the initial profile at large radii.
In runs A and B the profile for $r > 0.4R$ is well-fit by the $\beta$-model
with similar values of $\beta$. For $r < 0.4R$ the $\beta$-model produces
a slightly poorer fit.
As expected, the Hernquist model gives a poor fit for $r<0.2R$; but at
large radii both the $\beta$-model and the Hernquist model give good
fits, with the best-fit value of $\beta$ consistent with asymptotic slopes
of 4.0 and 3.8 for runs A and B, respectively, in agreement with the
results of Hernquist (1992, 1993) and Dubinski and Carlberg (1991).
For run C we were unable to obtain a statistically
reasonable fit with either function; the density profile in this case is
characterized by a large core ($\rho$ drops to 1/2 its central value near
$r=0.16R$), and while there is some evidence of power-law behavior outside
the core, the logarithmic slope appears to converge slowly.
In all cases the density profile is consistent with an
increased core radius. In runs A and B we see a large increase of the
central density of the merger remnant in comparison with the central densities
in the original clusters, while in run C the central density is slightly smaller than
the initial value.

These results suggest that collisions with modestly varying
impact parameter will produce merger remnants with substantially similar
density profiles varying as $r^{-4}$ (if infall is neglected)
at large radii, while the varying amount of shock heating
in such collisions will primarily manifest itself in modest differences in
the core radius. Most of the mass difference between the final merger remnant
and each of the initial clusters goes into a large increase in the central
density. A very large impact parameter, on the other hand, results in
a significantly larger core with a more complicated asymptotic profile.
In such cases the transfer of angular momentum out of the remnant core
by spiral shocks (as discussed in Section 4) may not be efficient enough
to produce a remnant whose structure is independent of $b_{\rm init}$.

\begin{figure}[tbh]
  \figurenum{15}
  \plotone{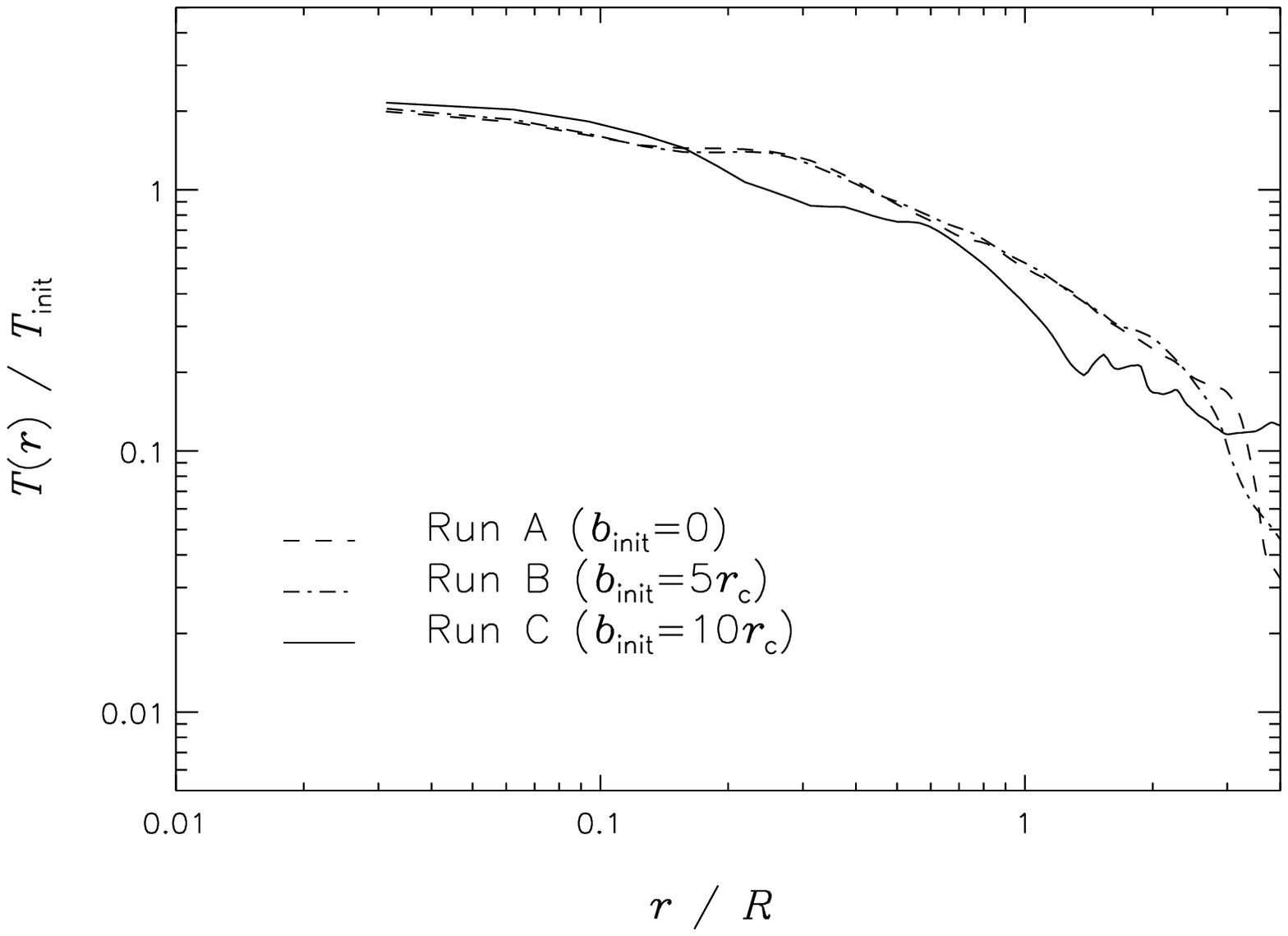}
  \caption{Angle-averaged temperature profiles, taken about the center of mass,
  for the merger remnants in runs A, B, and C at $t=12t_{\rm sc}$.}
  \label{Fig:Comparative Temperature Profile Plot}
\end{figure}

The angle-averaged temperature profiles for the three runs at
$t=12t_{\rm sc}$, plotted in Figure \ref{Fig:Comparative Temperature Profile
Plot}, are consistent with
these conclusions. For $r < 2R$ the temperature profiles
in runs A and B are virtually identical, with run C slightly hotter than the
other two for $r < 0.1R$ and substantially cooler outside this radius.
Inside $r = 0.2R$ the temperature in runs A and B is roughly
constant, dropping from
a central value of $2T_{\rm init}$ to about $1.5T_{\rm init}$. Outside this
radius the temperature drops approximately as $r^{-1}$. It is interesting
that the merger remnants in runs A and B are so similar; in the latter case
the gas inside $r = 0.3R$ is rotating about the center of mass at an azimuthal
velocity of about $0.5Rt_{\rm sc}^{-1}$, but rotation in this case does not
contribute much to its support against gravity.
For $b_{\rm init}=10r_c$, however, the rotation of the remnant has a very
significant effect on both the density and temperature profiles, suggesting
that if such extreme collisions occur at all, some evidence for them will be
present in the structure of the merger remnant.


\section{Conclusions}

We have used a new hydrodynamical code based on PPM and a multigrid isolated
potential solver to study the behavior of intracluster gas
in controlled collisions between equal-mass galaxy clusters at
different impact parameters. A summary of our findings follows.
Because of the lack of a collisionless component and of radiative cooling
in our calculations, as well as the simplified initial conditions, these
conclusions should not be applied to specific clusters or cosmological
models. Rather the physical insight suggested by these calculations
concerning the behavior of self-gravitating gas spheres should
be used to aid in the interpretation of more detailed calculations which
include dark matter and radiation or which use more realistic initial
conditions. The isolated potential solver we have developed for these
calculations is an important new component of a code we will use to perform
more realistic calculations in the near future.

Three main sets of shocks affect the progress of off-center mergers between
clusters of equal mass. A pair of roughly planar
shocks forms through the compression of the clusters' outer regions as
they approach one another. These shocks bound a disk-shaped outflow at a
varying angle to the collision axis. In off-center collisions, these planar
shocks become twisted into a spiral pattern through interaction with the
cluster cores. The ram pressure acting on each cluster determines the
size of the cluster's core. The cores drive a strong, roughly ellipsoidal
shock into the outer regions of the clusters, dissipating enough energy to
permit themselves to fall into
orbit about one another. In the wake of this shock the cluster cores
drive two curved bow shocks which gradually transfer the angular momentum of
the cores to the surrounding gas as they spiral inward. After the
cores collide, these spiral shocks continue to redistribute the angular
momentum and energy of the cores within the merger remnant. Material from
the outer regions, heated and driven outward by the ellipsoidal shock front,
then begins to fall back onto the merger remnant, creating a weak
accretion shock. The Mach numbers of all of the shocks are modest.
While the ellipsoidal and accretion shocks have been seen in previous
calculations of head-on mergers between clusters of galaxies
(e.~g.\ \cite{SchMul93}), we only
observe the spiral shocks in off-center mergers. We do not observe a
high-velocity, ordered flow like that seen in the calculations of
Roettiger \etal\ (1993). This is most likely because of the low initial velocity and
the lack of dark matter in our calculations.

We find that both
a physical criterion for equilibrium based on the virial theorem 
and the presence of readily observable bimodal or elliptical structure
in the projected X-ray surface brightness
and emission-weighted temperature
lead to equilibration times which are a significant fraction of the age of
the universe, given typical sound-crossing times of $\sim 1$~Gyr.
The virial equilibration time can be as large as 5--6 crossing times, while
isophotal structure is erased within 1--2 crossing times after the initial
core interaction for $b_{\rm init} = 0$ and $b_{\rm init} = 5r_c$.
For $b_{\rm init} = 10r_c$ the X-ray surface brightness shows evidence of
structure for several crossing times due to the protracted inspiral
of the cluster cores in this case. The head-on collision produces a much larger
increase in brightness during the collision than the other two cases, but it is
more short-lived; the luminosity enhancement due to the collision lasts
about one-half of a crossing time.

The angle-averaged density and temperature profiles of the merger remnant
show very little variation between a head-on collision and an off-center
collision with an impact parameter equal to five times the core radius.
The core radius increases slightly from its initial value due to shock
heating during
the collision, and the central density and temperature finish with values
about twice their initial values. The central $0.2R$ of the remnant is
roughly isothermal; outside this radius the temperature drops as $1/r$.
Outside the core the density drops steeply, with a $\beta$ value of
1.2 -- 1.3, corresponding to an asymptotic behavior closer to $r^{-4}$ than
to $r^{-3}$. Although the merger remnant rotates in the off-center case,
the remnant does not appear to be rotationally supported.
For an impact parameter of ten core radii, however, the remnant
is characterized by a large core, a low central density, and a complicated
profile at large radius, suggesting that the outward transfer of angular
momentum by spiral shocks is inefficient in this case.



\acknowledgments

The author would like to thank Don Lamb and Craig Sarazin for
advice and support during the completion of this work,
and Bruce Fryxell, Scott Dodelson, Kevin Olson, and Peter MacNeice for
useful and stimulating discussions. This work was primarily supported by
a NASA GSRP Fellowship (NGT-51322). Some work was also supported
under NASA NAG 5-3057. The calculations reported
here were performed at the Pittsburgh Supercomputing Center.








\clearpage

\nocite{Jam77}
\nocite{PreSch74}

\renewcommand{\and}{\ \&}
\newcommand{\JCP}{J.\ Comp.\ Phys.}
\newcommand{\MNRAS}{MNRAS}
\newcommand{\ApJ}{ApJ}
\newcommand{\ApJS}{ApJS}
\renewcommand{\AA}{A\&A}


\clearpage


\end{document}